\documentclass[11pt, a4paper]{article}
\pdfoutput=1

\usepackage{amsmath,amssymb}
\usepackage{mathrsfs,mathtools}
\usepackage{comment}
\usepackage{multirow}
\usepackage[utf8]{inputenc}
\usepackage{bm}
\usepackage{cite}
\usepackage{physics}
\usepackage{braket}
\usepackage{soul}
\usepackage[footnotesize]{caption}
\usepackage{acronym}

\usepackage{ifpdf}
\ifpdf
  \usepackage{graphicx, hyperref, xcolor}     
\else     
  \usepackage[dvipdfmx]{graphicx, hyperref, xcolor}     
 \fi
\usepackage{subcaption}

\usepackage{tikz}
\usepackage{tikz-feynman}
\tikzfeynmanset{compat=1.0.0}

\definecolor{rossoferrari}{HTML}{D9073D}
\definecolor{mediumblue}{HTML}{0000CD}
\definecolor{forestgreen}{HTML}{228B22}
\definecolor{desy_blue}{HTML}{009EE2}
\definecolor{desy_orange}{HTML}{FD8800}
\hypersetup{
setpagesize=false,
bookmarksnumbered=true,%
bookmarksopen=true,%
colorlinks=true,%
linkcolor=rossoferrari,
urlcolor=mediumblue,
citecolor=mediumblue,
linktocpage=false,
}

\usepackage[height=8.85in,width=6.8in]{geometry}

\usepackage{fourier}

\makeatletter
\@addtoreset{equation}{section}

\makeatother

\def\mcA{\mathcal{A}}

\def\mcP{\mathcal{P}}

\newcommand{\bme}{\bm{e}}
\newcommand{\bmJ}{\bm{J}}
\newcommand{\bmB}{\bm{B}}
\newcommand{\bmE}{\bm{E}}
\newcommand{\calO}{\mathcal{O}}
\newcommand{\bmx}{\bm{x}}

\definecolor{MONZA}{HTML}{CF000F}
\definecolor{DARKMAGENTA}{HTML}{8b008b}

\newcommand{\bae}[1]{\begin{align} #1 \end{align}}
\newcommand{\beae}[1]{\begin{equation}\begin{aligned} #1 \end{aligned}\end{equation}}
\newcommand{\bce}[1]{\begin{cases} #1 \end{cases}}
\newcommand{\dps}{\displaystyle}
\newcommand{\bfe}[4]{
\begin{figure}
	\centering
	\includegraphics[#1]{#2}
	\caption{#3}
	\label{#4}
\end{figure}}

\newcommand{\Uone}{\mathrm{U}(1)}
\newcommand{\DM}{\mathrm{DM}}
\newcommand{\app}{\mathrm{app}}
\newcommand{\num}{\mathrm{num}}

\begin{document}

\begin{titlepage}

\begin{center}

\hfill RESCEU-3/22\\
\hfill KEK-TH-2402\\

\vskip 1.in

{\Huge \bf \boldmath
Effective treatment of $\Uone$ gauge field and\\ 
charged particles in axion inflation\\
}

\vskip .8in

{\Large
Tomohiro Fujita,$^{\blacktriangledown,\triangledown}$
Jun'ya Kume,$^{\triangledown,\blacklozenge}$
Kyohei Mukaida,$^{\lozenge}$
and Yuichiro Tada$^{\bigstar}$
}

\vskip .3in

\begin{tabular}{ll}
$^\blacktriangledown$&\!\!\!\!\!\! \emph{Waseda Institute for Advanced Study, Shinjuku, Tokyo 169-8050, Japan}\\[.3em]
$^\triangledown$&\!\!\!\!\!\! \emph{Research Center for the Early Universe, University of Tokyo, Bunkyo, Tokyo 113-0033, Japan}\\[.3em]
$^\blacklozenge$&\!\!\!\!\!\! 
\emph{Department of Physics, Graduate School of Science, The University of Tokyo,} \\
&\!\!\!\!\!\! \emph{Hongo 7-3-1 Bunkyo-ku, Tokyo 113-0033, Japan}
\\[.3em]
$^\lozenge$&\!\!\!\!\!\! \emph{Theory Center, IPNS, KEK, 1-1 Oho, Tsukuba, Ibaraki 305-0801, Japan}\\[.3em]
$^\lozenge$&\!\!\!\!\!\! \emph{Graduate University for Advanced Studies (Sokendai), Tsukuba 305-0801, Japan}\\[.3em]
$^\bigstar$&\!\!\!\!\!\!
\emph{Institute for Advanced Research, Nagoya University,
Nagoya 464-8601, Japan}\\[.3em]
$^\bigstar$&\!\!\!\!\!\! \emph{Department of Physics, Nagoya University, Nagoya 464-8602, Japan}\\[.3em]
\end{tabular}

\end{center}
\vskip .6in

\begin{abstract}
\noindent
The axionic inflaton with the Chern--Simons coupling may generate $\Uone$ gauge fields and charged particles simultaneously.
In order to incorporate the backreaction from the charged particles on the gauge fields,
we develop a procedure to obtain an equilibrium solution for the gauge fields by treating the induced current as effective electric and magnetic conductivities. 
Introducing mean field approximation,
and numerically solving self-consistency equations, we find that the gauge field amplitudes are drastically suppressed. 
Interestingly, as the production becomes more efficient, 
the charged particles gain a larger part of the transferred energy from the inflaton and eventually dominate it.
Our formalism offers a basis to connect this class of inflationary models to a rich phenomenology such as baryogenesis and magnetogenesis.
\end{abstract}

\end{titlepage}

\renewcommand{\thepage}{\arabic{page}}
\setcounter{page}{1}

\tableofcontents

\newpage
\section{Introduction}

Among various models of inflation, the axion inflation is particularly well motivated because of the shift symmetry, which ensures the flatness of the inflaton potential~\cite{Freese:1990rb, Pajer:2013fsa}.
In order to reheat the universe after the inflation, however, one needs to introduce the coupling of the axion (inflaton) to the matter. Inclusion of the interaction term is acceptable if it also preserves the shift symmetry in this case. From such a perspective, the Chern--Simons (CS) coupling between the axion and gauge fields is naturally introduced since it can be rewritten as the derivative coupling of the axion field. 

If the CS coupling is present, gauge fields are generally sourced by the inflaton motion during the inflation. 
For Abelian gauge fields, it is known that there is a helicity dependent amplification of the magnetic fields due to the tachyonic instability~\cite{Turner:1987bw,Garretson:1992vt,Anber:2006xt,Fujita:2015iga,Adshead:2016iae,Cuissa:2018oiw}.
As a result, polarized gravitational waves are generated from the anisotropic shear-stress of the growing gauge field~\cite{Cook:2011hg,Barnaby:2011qe,Barnaby:2011vw,Anber:2012du,Namba:2015gja, Adshead:2019igv}. 
In addition, in the case of $\Uone_{Y}$ gauge field in the Standard Model (SM), such a substantial production of the cosmological Hyper-magnetic fields in the early universe has intriguing implications to particle cosmology. 
The intergalactic magnetic fields of $\Uone_{\text{em}}$ hinted by gamma-ray observations from distant blazars~\cite{Neronov:2010gir,Tavecchio:2010mk,Ando:2010rb,Dolag:2010ni,Essey:2010nd,Taylor:2011bn,Takahashi:2013lba,Finke:2015ona,Fermi-LAT:2018jdy,AlvesBatista:2020oio} can be originated from helical Hyper-magnetic fields~\cite{Caprini:2014mja,Fujita:2019pmi,Kamada:2020bmb}.
The present matter anti-matter asymmetry of the universe can be attributed to the helicity of the $\Uone$ gauge fields through the quantum anomaly~\cite{Giovannini:1997gp,Giovannini:1997eg,Fujita:2016igl,Kamada:2016eeb,Kamada:2016cnb,Kamada:2018tcs},
which can be originated from axion inflation~\cite{Anber:2015yca,Cado:2016kdp,Jimenez:2017cdr,Domcke:2019mnd,Domcke:2020quw}.
In such circumstances, giving a precise prediction for the gauge field production during inflation has significant importance.

However, the CS coupling to $\Uone_{Y}$ gauge field introduces complications because Hyper-charged SM particles are inevitably produced by the Schwinger effect of the sourced gauge field~\cite{Heisenberg:1936nmg,Schwinger:1951nm,Nikishov:1969tt,Bunkin:1969if}.
Consequently, one needs to take into account the backreaction from the created particles on the gauge fields.
Nevertheless, the effect of charged particles during inflation is mostly neglected in the above mentioned works except for a few recent attempts~\cite{Domcke:2018eki,Domcke:2019mnd,Domcke:2019qmm,Gorbar:2021rlt,Gorbar:2021zlr}. The evaluation of such a process is quite challenging due to the non-linearity of the system and the non-perturbative nature of the Schwinger effect. A systematic approximation of the gauge-field dynamics, or their spectrum, has not been fully established.
In order to deal with these difficulties and provide a reasonable prediction, we develop an effective treatment of the $\Uone$ gauge field and the charged particles.
Focusing on the dynamics in the middle of inflation, we assume constant $\dot{\phi}$ and Hubble parameter $H$ for simplicity in this work.

As for the charged particles, our strategy is basically similar to that in Ref.~\cite{Domcke:2018eki,Domcke:2019mnd} where the authors evaluated the induced current and identified the effect of fermions as the conductivity for the electromagnetic fields. Once the induced current is expressed in terms of the gauge field, the equation of motion for the gauge field becomes non-linear. In order to obtain the ansatz for the field dynamics, we develop a mean field approximation for the gauge field which effectively reduces the equation of motion to the linear equation. Roughly speaking, a single $k$-mode of interest, which is going to grow, is identified as the perturbation. The other modes including the dominant mode, which behave as the static and homogeneous background, are treated as the mean field.
We are forced to introduce both the electric and magnetic conductivity parameters of the Schwinger current for the perturbation, which depend on the mean field strength.\footnote{\samepage
  This is in contrast to Refs.~\cite{Domcke:2018eki,Domcke:2019mnd} and \cite{Sobol:2019xls,Gorbar:2021rlt,Gorbar:2021zlr}.
  There, either the magnetic or electric conductivity is introduced respectively, but not simultaneously.
  See also Eqs.~\eqref{Sigma1} to \eqref{SigmaE_and_SigmaB}.
}
Once the single mode starts to evolve, it becomes dominant at some point and in turn behaves as part of the mean field with respect to the subsequently generated mode. This requires the self-consistent condition that the solution reproduces the mean field strength or the certain value of the conductivity parameters. By numerically computing the self-consistent value of the Schwinger conductivity, we obtain the equilibrium solution of the gauge field where the back-reaction from the charged particles is effectively taken into account. We also carefully check the validity of our expression for the induced current and the mean field approximation.

The rest of the paper is organized as follows. In Sec.~\ref{The U(1) gauge field without charged particles}, we make a brief review of the $\Uone$ gauge field production in axion inflation with the CS coupling. 
Then we investigate the effect of the charged particles in Sec.~\ref{Sec: charged particles as the conductivity}, through the mean field approximation and the self-consistent Schwinger conductivity. To simplify the analysis, single species of a charged particle is concerned instead of the full SM. In order to discuss the validity of our approximation and investigate the self-consistent solutions, we perform several consistency checks in Sec.~\ref{consistency check}. Sec.~\ref{conclusion} is devoted to the conclusion.

\section{\boldmath The $\Uone$ gauge field without charged particles}
\label{The U(1) gauge field without charged particles}

In this section, we briefly review the evolution of the $\Uone$ gauge field without the charged particles. We consider the following Lagrangian in which the inflaton $\phi$ is coupled with the $\Uone$ gauge field $A_\mu$ through the CS coupling;
\begin{align}
\mathcal{L}=\frac{1}{2}\partial_\mu \phi\partial^\mu \phi-V(\phi)
-\frac{1}{4}F_{\mu\nu}F^{\mu\nu}-\frac{1}{4f}\phi F_{\mu\nu}\tilde{F}^{\mu\nu},
\end{align}
where $F_{\mu\nu}\equiv \partial_\mu A_\nu - \partial_\nu A_\mu$ is the electromagnetic field strength and $\tilde{F}^{\mu\nu}\equiv \epsilon^{\mu\nu\rho\sigma}F_{\rho\sigma}/(2\sqrt{-g})$ is its dual. The determinant of the spacetime metric is denoted by $g$ and the totally anti-symmetric tensor is defined by $\epsilon^{0123}=1$.
In this paper, we do not specify the inflaton potential $V(\phi)$ or the value of the axion decay constant $f$.
In the spatially flat FLRW universe, $\dd s^2= a^2(\tau)(\dd \tau^2-\dd \bm{x}^2)$, and the Coulomb gauge in vacuum, $A_0=\partial_iA_i=0$,
EoM for the comoving gauge field is given by
\begin{align}
\partial_\tau^2 A_i-\partial_j^2 A_i-\frac{1}{f} (\partial_\tau \phi)
\epsilon_{ijl}\partial_j A_l=0,
\label{Original A EoM}
\end{align}
where the conformal time is denoted by $\tau$
and the rank-$3$ totally anti-symmetric tensor is $\epsilon_{123} = 1$.
Note that we ignore the perturbation of the inflaton $\phi$ and consider it as a function of time.
The gauge field is decomposed by the circular polarization and quantized as
\begin{align}
 A_i(\tau, \bm{x})
 &=
 \sum_{\lambda=\pm} \int \frac{{\rm d}^3 k}{(2\pi)^3}
 e^{i \bm{k \cdot x}} e_{i}^{(\lambda)}(\hat{\bm{k}})
 \hat{A}_\lambda(\tau,\bm k),
 \\
 \hat{A}_\lambda(\tau,\bm k)
 &= \hat{a}_{\bm{k}}^{(\lambda)} \mcA_\lambda(\tau,k)  + \hat{a}_{-\bm{k}}^{(\lambda) \dag} \mcA_\lambda^*(\tau,k),
\label{quantization}
\end{align}
where $e^{(\pm)}_i(\hat{\bm{k}})$ are the right/left-handed polarization vectors which satisfy $i \bm{k} \cp \bm e^{(\pm)}(\hat{\bm{k}})=\pm k\, \bm{e}^{(\pm)}(\hat{\bm{k}})$, and
$\hat{a}_{\bm{k}}^{(\pm) \dag}$/$\hat{a}_{\bm{k}}^{(\pm)}$ are the creation/annihilation operators which satisfy the usual commutation relation, $[\hat{a}^{(\lambda)}_{\bm{k}},\hat{a}^{(\sigma) \dag}_{-\bm{k}'}]
= (2\pi)^3\delta(\bm{k}+\bm{k}')\delta^{\lambda \sigma}$.

During inflation $aH=-1/\tau$, the EoM for the mode function is written as
\begin{align}
\left[ \partial_\tau^2 +k^2 \pm 2k \frac{\xi}{\tau} \right] \mcA_\pm(\tau,k)=0,
\label{EoMforA}
\end{align}
with
\begin{align}
\xi\equiv \frac{\partial_\tau \phi}{2f aH}=\frac{\dot{\phi}}{2fH},
\end{align}
where dot denotes the cosmic time derivative.
If $\xi>0$, for instance, $\mcA_+$ modes undergo an exponential enhancement
around the horizon crossing.
With Bunch--Davies vacuum and constant $\xi$, one can find the analytic solution for $\mcA_+$ as
\begin{align}
\mcA_+(\tau,k)=\frac{1}{\sqrt{2k}}e^{\pi\xi/2}W_{- i\xi,1/2}(2ik\tau),
\label{A_sol}
\end{align}
where $W_{\alpha,\beta}(z)$ is the Whittaker $W$ function.
For brevity, we define this Whittaker function and its derivative as
\begin{align}
W(z) \equiv W_{-i\xi,1/2}(-2i z),
\qquad
W'(z)\equiv \partial_z W_{-i\xi,1/2}(-2iz).
\end{align}
With this solution, the {\it physical} electromagnetic spectra for the $+$ mode are given by
\begin{align}
\tilde{\mcP}_{BB}^+(\tau,k)&=
a^{-4}\mcP_{BB}^+(\tau,k)=
\frac{k^5}{2\pi^2 a^4}\left|\mcA_+(\tau,k)\right|^{2}
=\frac{|k\tau|^4 H^4}{4\pi^2}e^{\pi\xi} \left|W(-k\tau)\right|^2,
\label{PB tilde}
\\
\tilde{\mcP}_{EE}^+(\tau,k)&=
a^{-4}\mcP_{EE}^+(\tau,k) = \frac{k^3}{2\pi^2 a^4} \left|\partial_\tau \mcA_+(\tau,k)\right|^{2}
=\frac{|k\tau|^4 H^4}{4\pi^2 }e^{\pi\xi} \left|W'(-k\tau)\right|^2,
\label{PE tilde}
\\
\tilde{\mcP}_{BE}^+(\tau,k)&=
a^{-4}\mcP_{BE}^+(\tau,k) = -\frac{k^4}{2\pi^2 a^4} \mcA_+(\tau,k)\partial_\tau \mcA_+^*(\tau,k)
=\frac{|k\tau|^4 H^4}{4\pi^2 }e^{\pi\xi} W(-k\tau)W'^*(-k\tau),
\end{align}
where $\mcP_{XX}^\lambda$ are the {\it comoving} spectra and $\mcP_{EB}^\lambda=(\mcP_{BE}^{\lambda})^*$.
%
\begin{figure}[tbp]
  \begin{center}
  \includegraphics[width=100mm]{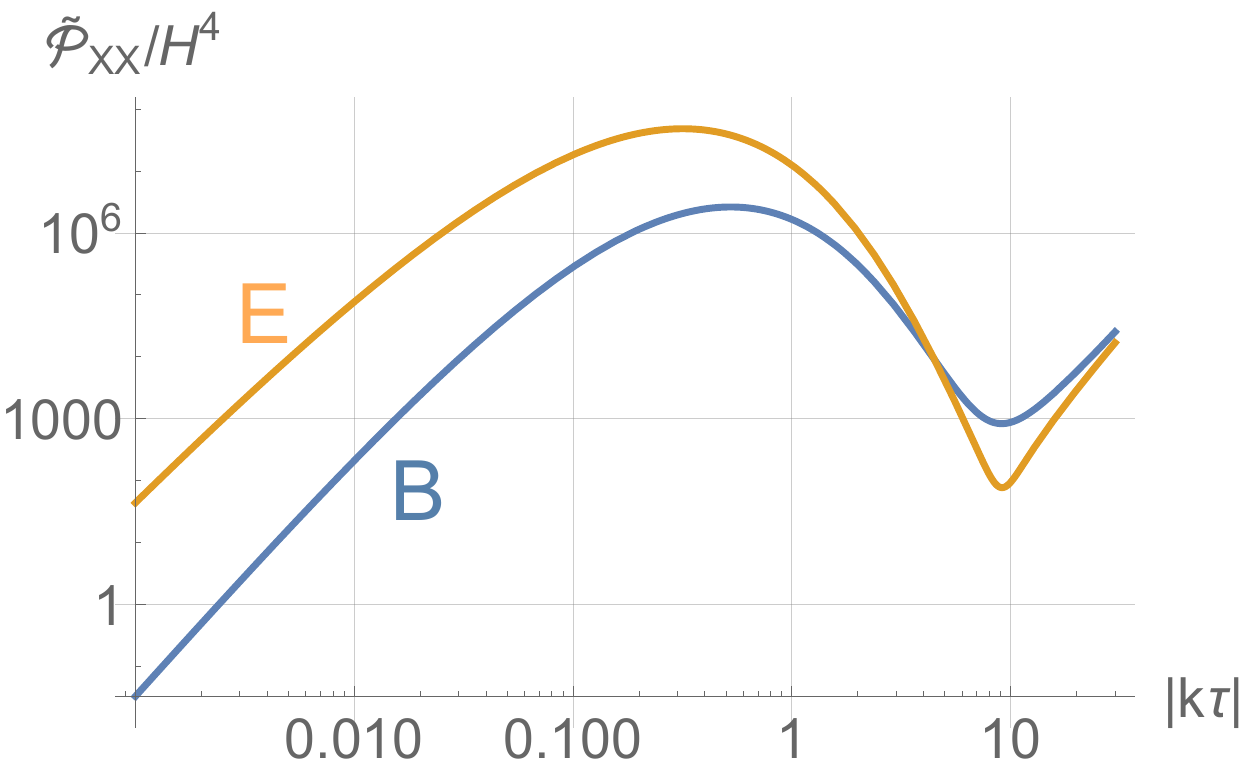}    
  \end{center}
  \caption{The electromagnetic power spectra, $H^{-4}\tilde{\mcP}_{BB}^+$ (blue) and $H^{-4}\tilde{\mcP}_{EE}^+$ (orange) for $\xi=5$, given in Eqs.~\eqref{PB tilde} and \eqref{PE tilde}. Their peak amplitudes are much larger than the Hubble scale, while the peak length scale and the damping time scale are roughly given by the Hubble scale.
  $\tilde{\mcP}_{EE}$ is larger than $\tilde{\mcP}_{BB}$ by $\mathcal{O}(\xi^2)$.}
  \label{x510}
\end{figure}
%
Fig.~\ref{x510} shows the auto-power spectra, $\tilde{\mcP}_{EE}^+$ and $\tilde{\mcP}_{BB}^+$,
in the case of $\xi=5$.
One observes that $\tilde{\mcP}_{EE}^+$ is larger than $\tilde{\mcP}_{BB}^+$
and their peak amplitudes greatly exceeds the Hubble scale.
Namely, the {\it physical} electromagnetic fields, 
$\tilde{E}=a^{-2}E, \tilde{B}=a^{-2}B$, have a hierarchy,
\begin{equation}
    \tilde{E} > \tilde{B} \gg H^2.
\end{equation}
Note here that one may show $\tilde E \sim \xi \tilde{B}$ for $\xi \gg 1$, which leads to the first inequality.
The power spectra reach their peak amplitudes at $|k\tau|\simeq \xi^{-1}$
and thus the correlation length scale is roughly estimated as
\begin{equation}
    L_\mathrm{em} \simeq \frac{\xi}{H}.
\end{equation}
After passing their peaks, the physical spectra decay as $\sim a^{-4}$,
and their dynamical time scale is given by \begin{equation}
    t_\mathrm{em} \simeq \frac{1}{4H}.
\end{equation}
The net electromagnetic field values can fluctuate within this time scale as these modes are continually generated and diluted, but on average, they are understood to be in stable equilibrium.
One can also show that the phase rotations of $W(-k\tau)$ and $W'(-k\tau)$ stop
at around $|k\tau|=2\xi$, and their terminal phases always satisfy $W(-k\tau)W'^*(-k\tau)=W^*(-k\tau)W'(-k\tau)=-|W(-k\tau)||W'^*(-k\tau)|$.
Using it, we obtain
\begin{equation}
    \frac{\tilde{\mcP}_{BE}^+}{\sqrt{\tilde{\mcP}_{EE}^+\tilde{\mcP}_{BB}^+}}
    \xrightarrow{|k\tau|\ll 2\xi}  -1.
    \label{Antiparalel}
\end{equation}
This relation implies that the produced electric and magnetic fields take the anti-parallel
configuration, $\hat{\bm E}\cdot \hat{\bm B}=-1$, where $\hat{\bmE}$ and $\hat{\bmB}$ are the unit vectors $\bmE/\abs{\bmE}$ and $\bmB/\abs{\bmB}$.
This is a manifestation of the parity violating nature of the axion and the CS coupling,
while the minus sign here is merely the consequence of our choice, $\xi>0$.
For $\xi<0$, the electromagnetic fields would be parallel.
Note that this relation becomes a good approximation soon after $\abs{k\tau}=2\xi$, particularly including the peak mode $\abs{k\tau}\sim1/\xi$. Therefore, the electromagnetic fields can be safely assumed to be anti-parallel (or parallel for $\xi<0$) in average.

In summary, we have observed the following four properties of the electromagnetic fields sourced by the axionic inflaton in the case without the charged particles:
(i) for $\xi\gtrsim \mathcal{O}(1)$ of our interest,
strong electromagnetic fields are produced: $\tilde{E}, \tilde{B}\gg H^2$,
(ii) the correlation length scale is $L_\mathrm{em}\sim 1/H$,
(iii) the dynamical time scale is $t_\mathrm{em}\sim H^{-1}$, and
(iv) the electric and magnetic fields are anti-parallel, $\hat{\bm E}\cdot \hat{\bm B}\simeq -1$, for $\xi > 0$.

\section{Charged particles as the conductivity}
\label{Sec: charged particles as the conductivity}

\subsection{Schwinger current}

In this section, we develop an effective method to take charged particles into consideration. 
We first introduce a matter field charged under the $\Uone$ gauge symmetry.
To be concrete, let us consider a Dirac fermion which has a unit charge $e$ and mass $m$.
Our interaction term between the $\Uone$ gauge field $A_\mu$ and the Dirac fermion $\psi$ is
\begin{equation}
\sqrt{-g}\mathcal{L}_{\rm int}= -a^4 e J^\mu A_\mu,
\qquad
J^\mu=\bar{\psi}\gamma^\mu \psi,
\label{int lagrangian}
\end{equation}
where the gamma matrix satisfies $\{\gamma^\mu,\gamma^\nu\}=g^{\mu\nu}$.
Under an environment of strong electromagnetic fields, the charged fermions are generated through the Schwinger effect and they backreact onto the gauge field EoM as an induced current.
We consider the small mass regime, $m^2 \ll e \tilde E$, where the Schwinger production and the backreaction are the most efficient.

Let us assume the charge neutrality $J_0=0$ and the current conservation $\partial_i J_i=0$ on the scales of our interest. Then we can formally incorporate the above interaction into the EoM for $A_i$~\eqref{Original A EoM} as
\begin{equation}
    \partial_\tau^2 A_i-\partial_j^2 A_i+\frac{2\xi}{\tau} \epsilon_{ijl}\partial_j A_l
    =a^2 e J_i.
    \label{Ai EoM w/ full current}
\end{equation}
It is generally difficult to perform a precise calculation of the induced current because one has to solve the coupled dynamics of the gauge field and the Dirac field.
However, one can solve the fermion dynamics under the assumption of homogeneous, static, and anti-parallel {\it physical} electromagnetic field, and estimate the production of the induced current as~\cite{Domcke:2018eki}\footnote{
  Eq.~(4.14) of Ref.~\cite{Domcke:2018eki} says $e J_i^{\DM}\propto a^{3}$ instead of $e J_i \propto a$ in our Eq.~\eqref{Schwinger conductivity}.
  This apparent discrepancy comes from the differences in definition.
  In Ref.~\cite{Domcke:2018eki}, $J_i^\text{DM}$ is a comoving current which is defined after rescaling fermion fields appropriately.
  However, our $J_i$ is just a current of a lower index without rescaling fermions.
  This implies $J^\text{DM}_i = - J^i_\text{DM} = - a^4 J^i = a^2 J_i$.
}
\begin{align}
    \partial_\tau (a^2 e J_i) = \frac{e^3BE_i}{2\pi^2}\coth\pqty{\frac{\pi B}{E}}.
    \label{Dcurrent}
\end{align}
Note that this expression remains valid when scattering among the produced fermions is negligible compared to the acceleration due to the Schwinger effect. Hereafter we assume that this is the case for our system and a qualitative discussion is summarized in Appendix~\ref{Thermalization}. More essentially,
$E/a^2=\tilde{E}\simeq{\rm const.}, B/a^2=\tilde{B}\simeq{\rm const.,}$ and $\bm E\cdot \bm B=-EB$ are assumed in Eq.~\eqref{Dcurrent}.
Although these assumptions are not exactly fulfilled, the properties of the electromagnetic fields (i)--(iv)
summarized at the end of Sec.~\ref{The U(1) gauge field without charged particles} imply that the above expression is the leading approximation of the induced current in our system. A typical comoving wave length of the Dirac fermion right after the Schwinger production is $L_\psi \sim 1/\sqrt{eE}$, 
and a typical time scale of the Schwinger production is also $t_\psi\sim 1/\sqrt{eE}$.
The properties of the electromagnetic fields (i)--(iii)
indicate that we have a significant scale separation between the fermions and the electromagnetic fields,
\begin{equation}
    L_\psi \sim t_\psi \sim (eE)^{-1/2} \ll
    L_\mathrm{em} \sim t_\mathrm{em} \sim H^{-1}.
    \label{scaleseparation}
\end{equation}
Thus the Dirac fermions see the electromagnetic fields as a static and homogeneous background.
For the time being, we also assume the anti-parallelism of the electromagnetic fields which is apparently ensured by the property (iv). There is a subtlety and we will address it in the next subsection. One might worry that the backreaction of the Dirac fermions could significantly change these properties of the electromagnetic fields. 
We will confirm its validity at the end of this section.

Now the Dirac fermions are effectively integrated out and the coupled equations,
\eqref{Ai EoM w/ full current} and \eqref{Dcurrent}, only in terms of the gauge field describe our system.
However, they are still hard to be solved as they are non-linear equations in $A_i$ due to the Schwinger current~\eqref{Dcurrent}.
In order to solve them, we then introduce two approximations; constant physical electromagnetic fields and mean field approximations.
The first approximation is that the physical electromagnetic fields are static, $\tilde{E},\tilde{B}\simeq{\rm const.}$, for a longer time scale than the Hubble time, $H^{-1}$.
This approximation allows us to integrate Eq.~\eqref{Dcurrent} and obtain
\begin{equation}
    eJ_i \simeq \frac{e^3 B E_i}{6\pi^2 a^3 H} \coth\left(\frac{\pi B}{E}\right).
\label{Schwinger conductivity}
\end{equation}
Since the correlation time of the electromagnetic fields is about the Hubble time, $t_\mathrm{em}\sim H^{-1}$, this approximation may not be very accurate. Nevertheless, our entire system is stationary which is driven only by the constant energy injection from the inflaton, and we expect that the time-averaged physical amplitude of the electromagnetic fields is constant. Moreover, $t_\mathrm{em}\sim H^{-1}$ implies that we would have at most an $\mathcal{O}(1)$ correction to Eq.~\eqref{Schwinger conductivity} because no shorter time scale is involved.
We will check the validity of this approximation in Sec.~\ref{pert_current}.

\subsection{Mean field approximation}

By substituting Eq.~\eqref{Schwinger conductivity} into the right hand side of Eq.~\eqref{Ai EoM w/ full current}, the coupled equations are reduced into a single equation of motion describing our system.
However, the resultant EoM is still non-linear and it is generally difficult to be solved. In order to seek the possible form of the solution, we here utilize the mean field approximation for the gauge field which can reduce the EoM to the linear equation. 
Let us decompose the electromagnetic fields into a mean part and a perturbation part as
\begin{equation}
    \bm{E}(\tau,\bm x) \simeq \bm{E}_0 + \delta \bm{E}(\tau,\bm x),
    \qquad
    \bm{B}(\tau,\bm x) \simeq \bm{B}_0 + \delta \bm{B}(\tau,\bm x).\label{decomposition}
\end{equation}
Here, we identify the perturbation part $\delta \bm{E}(\tau,\bm x)$ and $\delta \bm{B}(\tau,\bm x)$ as the contribution from a single $\bm{k}$-mode in Fourier space and the mean part $\bm{E}_0$ and $\bm{B}_0$ as the summation of all the other modes.\footnote{In this sense, the mean part is implicitly labeled by $\bm{k}$. Here we assume that any single mode is subdominant compared to the mean part and thus mean field approximately take common value $\bm{E}_0$ and $\bm{B}_0$ for all $\bm{k}$.} 
This decomposition enables us to consider the evolution of the single mode $\delta A_i(\tau,\bm k)$ (perturbation) as a function of the static physical background $\tilde{\bm{E}}_0$ and $\tilde{\bm{B}}_0$ (mean field).
The self-consistency condition is required by the assumption that the background fields themselves consist of the summation of these perturbations. That is, we assume that the amplitudes of the background fields are given by
\bae{\label{eq: consistency condition}
    \tilde{E}_0=\sqrt{2\rho_E(\tilde{E}_0,\tilde{B}_0)} \qc 
    \tilde{B}_0=\sqrt{2\rho_B(\tilde{E}_0,\tilde{B}_0)},
}
on average, where $\rho_E$ and $\rho_B$ denote the physical energy densities of the perturbations $\delta\bm{E}$ and $\delta\bm{B}$ (see Eq.~\eqref{eq: rhoE and rhoB}). The consistent background amplitudes $\tilde{E}_0$ and $\tilde{B}_0$ will be found so that the input $\tilde{E}_0$ and $\tilde{B}_0$ return the same values through the condition~\eqref{eq: consistency condition}.
Below, we further assume that the background fields are anti-parallel, $\bm{E}_0\cdot\bm{B}_0=-E_0B_0$, because they are mainly contributed by sufficiently superhorizon modes $\abs{k\tau}\ll2\xi$.
In such a way, we will approximately solve the fully non-linear coupled equations~\eqref{Ai EoM w/ full current} and \eqref{Dcurrent}.
The validity of these approximations on the mean field will be checked in Sec.~\ref{consistency check}.

In order for the perturbative prescription, we further have to decompose the Schwinger current.
As already mentioned, the expression for the induced current \eqref{Schwinger conductivity} is only applicable to the static, homogeneous, and anti-parallel electromagnetic fields. Thanks to the scale separation \eqref{scaleseparation}, the contributions to the electromagnetic fields even from the modes under the tachyonic instability on sub-horizon scales seem to be sufficiently static and homogeneous for the fermions. 
Based on this observation, Ref.~\cite{Domcke:2018eki} uses the Schwinger current~\eqref{Schwinger conductivity} even for the perturbation part and solve the perturbation equation self-consistently by regarding it as a magnetic conductivity, $e \delta\bm{J} = \sigma_B \delta \bm{B}$.
Refs.~\cite{Gorbar:2021zlr,Gorbar:2021rlt} instead identify the Schwinger current~\eqref{Schwinger conductivity} for the perturbation as an electric conductivity, $\delta \bm{J} = \sigma_E \delta \bm{E}$.

However, as seen in Eq.~\eqref{Antiparalel}, the contributions from such modes at $|k\tau|\sim 2\xi$ are not necessarily anti-parallel. In other words, the perturbation part, $\delta \bm E$ and $\delta \bm B$,  
may not be anti-parallel, while we expect
that the background part, which is dominantly contributed by longer modes $|k\tau|\ll 2\xi$, approximately takes the anti-parallel configuration. 
Thus, we need to generalize the expression for the induced current in Eq.~\eqref{Schwinger conductivity} by relaxing the anti-parallel condition of the perturbation part.
As we will see soon, this forces us to introduce the both electric and magnetic conductivities simultaneously in the perturbed equation of motion.

As is well known, one can always find a coordinate frame where the electromagnetic fields are anti-parallel and Eq.~\eqref{Schwinger conductivity} is valid. Then one obtains the expression for the induced current in the original frame through the Lorentz boost.
As we describe in Appendix~\ref{Lorentz boost}, at the leading order of perturbation in $\epsilon \sim \delta E/E_0\sim\delta B/B_0$, it turns out that the amplitude of the full current is simply expressed by Eq.~\eqref{Schwinger conductivity} with $E(\tau, \bm{x})$ and $B(\tau, \bm{x})$, but its direction is given by
\bae{
    \hat{\bmJ}=\bqty{1-\frac{E_0\delta E_z-B_0\delta B_z}{E_0^2+B_0^2}}\bme_z+\frac{E_0\delta\bmE-B_0\delta\bmB}{E_0^2+B_0^2},
}
where we took the direction of the background electric field as $z$-direcion without loss of generality and
hat denotes the normal vector, $\hat{\bm J}=\bm{J}/|\bm{J}|$.
As a result, the induced current can be expanded as follows:
\beae{
    a^2e\bmJ&=a^2e(\bmJ_0+\delta\bmJ), \\
    a^2e\bmJ_0&=\frac{e^3B_0E_0}{6\pi^2aH}\coth\pqty{\frac{\pi B_0}{E_0}}\bme_z, \\
    a^2e\delta\bmJ&=
    \frac{e^3}{6\pi^2aH}\left[\pqty{\frac{B_0^3\delta E_z-E_0^3\delta B_z}{E_0^2+B_0^2}\coth\pqty{\frac{\pi B_0}{E_0}}+(B_0\delta E_z+E_0\delta B_z)\frac{\pi B_0}{E_0}\csch^2\pqty{\frac{\pi B_0}{E_0}}}\bme_z \right. \\
    &\qquad\left.+\frac{E_0^2B_0\delta\bmE-B_0^2E_0\delta\bmB}{E_0^2+B_0^2}\coth\pqty{\frac{\pi B_0}{E_0}}\right].
    \label{dJ in real space}
}
Here $\bmJ_0$ is the contribution from the mean field which is balanced with the mean part itself. 
Since the induced current is now expressed in the linear order of $\delta A_i(\tau,\bmx)$, the EoM is approximately reduced to the linear equation and we can find the solution for the $\delta A_i$. 
Note again that these expressions for $\bmJ_0$ and $\delta\bmJ$ assume the static physical electromagnetic fields for both background and perturbation $\tilde{E}_0, \tilde{B}_0, \delta\tilde{E}, \delta\tilde{B} \simeq{\rm const.}$
as well as the antiparallelism of the background fields, $\hat{\bm E}_0\cdot \hat{\bm B}_0\simeq -1$
before solving the linearized EoM. We will confirm the validity of this treatment in Sec.~\ref{pert_current}.

\subsection{Self-consistent evolution}\label{self-consistent evolution}

We are ready to study how the backreaction of the induced current affects the evolution of the electromagnetic fields. Since we obtained the expression of the Schwinger current in the mean-field approximation, let us plug it into the perturbed version of Eq.~\eqref{Ai EoM w/ full current}. Moving to the Fourier space, we find the EoMs for the mode function $\mcA_+^{(\sigma)}$ including the effect of the induced current (see Appendix~\ref{Derivation of the perturbed EoM for the gauge field} for derivation) 
\begin{align}\label{EoMforA+w/sigma}
    \left[ \partial_\tau^2-\frac{\Sigma_E+\Sigma_{E^{\prime}}\sin^2\theta_{\bm k}}{\tau} \partial_\tau +k^2 + \frac{k}{\tau}\left(2\xi-(\Sigma_B+\Sigma_{B^{\prime}} \sin^2\theta_{\bm k}) \right) \right] \mcA_+^{(\sigma)}(\tau,\bm k)&=0,
\end{align}
where $|\mcA_+^{(\sigma)}| \gg |\mcA_-^{(\sigma)}|$ is assumed. 
Here $\theta_{\bm k}$ is the angle between $\hat{\bm E}_0$ and $\hat{\bm k}$, and then we have $\hat{\bm E}_0 \cdot \bm{e}^\pm (\hat{\bm k})=-\sin\theta_{\bm k}/\sqrt{2}$.\footnote{Without loss of generality, we can take the $z$-axis in parallel with $\bm{E}_0$. A polarization vector with $\bm k$ pointing to $(\theta,\varphi)$ is given by $\bm{e}^\pm(\hat{\bm k})=(\cos \varphi \cos\theta\mp i \sin \varphi,\ \sin \phi \cos\theta\pm i \cos \varphi,\  -\sin\theta)/\sqrt{2}$.
Thus, one finds $\hat{\bm E}_{0}\cdot {\bm e}^\pm(\hat{\bm k})=-\sin\theta/\sqrt{2}$. \label{footnote: polarizaion vector}}\,
The superscript $(\sigma)$ indicates that we include the conductivity parametrized by
\begin{align}
    \Sigma_{E} &\equiv \frac{e^3B_0}{6\pi^2a^2H^2}\left(\frac{E_0^2}{E_0^2+B_0^2}\coth\pqty{\frac{\pi B_0}{E_0}}\right),
    \label{Sigma1}\\
    \Sigma_{B} &\equiv \frac{e^3E_0}{6\pi^2a^2H^2}\left(\frac{B_0^2}{E_0^2+B_0^2}\coth\pqty{\frac{\pi B_0}{E_0}}\right),
    \label{Sigma2}\\
    \Sigma_{E^{\prime}} &\equiv \frac{e^3B_0}{12\pi^2a^2H^2}\pqty{\frac{B_0^2}{E_0^2+B_0^2}\coth\pqty{\frac{\pi B_0}{E_0}}+\frac{\pi B_0}{E_0}\csch^2\pqty{\frac{\pi B_0}{E_0}}},
    \label{Sigma3}\\
    \Sigma_{B^{\prime}} &\equiv \frac{e^3E_0}{12\pi^2a^2H^2}\pqty{\frac{E_0^2}{E_0^2+B_0^2}\coth\pqty{\frac{\pi B_0}{E_0}}-\frac{\pi B_0}{E_0}\csch^2\pqty{\frac{\pi B_0}{E_0}}}.\label{Sigma4}
\end{align}
When the {\it physical} mean electromagnetic fields are static, 
$\tilde{E}_0=E_0/a^2=$const. and $\tilde{B}_0=B_0/a^2=$const.,
these conductivity parameters are also constant.

Let us solve Eq.~\eqref{EoMforA+w/sigma}.
Changing the time variable into $z\equiv -k\tau$, the EoM reads
\begin{align}\label{eq: A+sigma in x}
\left[ \partial_z^2-\frac{\Sigma}{z} \partial_z +1 - \frac{2\xi_\mathrm{eff}}{z} \right] \mcA_+^{(\sigma)}=0,
\end{align}
with
\begin{align}
\label{SigmaE_and_SigmaB}
\Sigma\equiv\Sigma_E+\Sigma_{E^{\prime}}\sin^2\theta_{\bm k},
\qquad
\xi_\mathrm{eff}\equiv \xi-\frac{1}{2}\left(\Sigma_B+\Sigma_{B^{\prime}}\sin^2\theta_{\bm k}\right).
\end{align}
This equation clearly shows the need of both the electric conductivity $\Sigma$ and the magnetic conductivity $\Sigma_{B(B^{\prime})}$.
The effect of non-zero electric conductivity $\Sigma$ appear as a friction term, while the magnetic conductivity $\Sigma_{B(B^{\prime})}$ reduces $\xi$ effectively.\footnote{The minus sign of the friction term originates in the time variable $x$ which is positive but decreases as time goes on. If one changes the time variable from $x$ into the cosmic time $t$, the friction term would have a positive sign.}
The general solution of the EoM is found as
\begin{align}
\mcA_+^{(\sigma)}(\tau,\bm k)=
\frac{1}{\sqrt{2k}}e^{\pi\xi_\mathrm{eff}/2} z^{\Sigma/2}\Big[ c_1 W_{-i\xi_\mathrm{eff},(\Sigma+1)/2}(-2iz)+c_2 M_{-i\xi_\mathrm{eff},(\Sigma+1)/2}(-2iz) \Big],
\label{Asigma}
\end{align}
where $c_1$ and $c_2$ are integration constants and $M_{\alpha,\beta}(z)$ is the Whittaker $M$ function.
In the sub-horizon limit, the scale of the electromagnetic fields becomes smaller than that of the fermions and our approximation based on Eq.~\eqref{scaleseparation} is no longer valid. 
Moreover, the induced current should vanish in deep subhorizon because a deep UV mode is not influenced by any environment involving IR modifications.
Thus, at some early time $\tau=-\gamma/k$ parametrized by $\gamma$,
we connect the above solution 
to the original solution without the conductivity Eq.~\eqref{A_sol} so that it corresponds to the vacuum in deep subhorizon.\footnote{Without such a prescription, the friction term even decreases the vacuum fluctuation which would not be acceptable. Furthermore, this suppression depends on the initial time of inflation (see Refs.~\cite{Gorbar:2021rlt,Gorbar:2021zlr}). Our prescription successfully avoids these problems. \label{footnote}}
The integration constants are obtained as
\begin{align}
    c_1 &=ie^{\frac{\pi}{2}(\xi-\xi_\mathrm{eff})}\gamma^{-\frac{\Sigma}{2}}\frac{\Gamma(1+i\xi_\mathrm{eff}+\Sigma/2)}{2\Gamma(\Sigma+2)}
    \left[W(\gamma) M_\Sigma'(\gamma) - W'(\gamma) M_\Sigma(\gamma)+\frac{\Sigma}{2\gamma}W(\gamma) M_\Sigma(\gamma) \right],
    \\
    c_2 &=-ie^{\frac{\pi}{2}(\xi-\xi_\mathrm{eff})}\gamma^{-\frac{\Sigma}{2}}\frac{\Gamma(1+i\xi_\mathrm{eff}+\Sigma/2)}{2 \Gamma(\Sigma+2)}
    \left[W(\gamma) W_\Sigma'(\gamma)-W'(\gamma) W_\Sigma(\gamma)+\frac{\Sigma}{2\gamma}W(\gamma) W_\Sigma(\gamma) \right],
\end{align}
where
$W_\Sigma(z)\equiv W_{-i\xi_\mathrm{eff},(1+\Sigma)/2}(-2iz),\ 
M_\Sigma(z)\equiv M_{-i\xi_\mathrm{eff},(1+\Sigma)/2}(-2iz)$,
and $X'\equiv \partial_z X(z)$ for $X=W_\Sigma$ and $M_\Sigma$.
One might wonder if this connection should be at around the fermion scale $E_{\psi} \sim \sqrt{eE}$, which is higher than $2\xi H$. In such a regime, however, one cannot apply the expression of induced current~\eqref{Schwinger conductivity} obtained by integrating out the fermions. 
In the following, we set $\gamma=2\xi$ so that the electromagnetic fields start to be affected by the fermion backreaction immediately after they leave the vacuum state.
We will see that these two scales are not so different practically.

Using the mode function \eqref{Asigma}, we write the power spectra as
\begin{align}
\tilde{\mcP}_{BB}^{+(\sigma)}(z,\theta_{\bm k})&=
\frac{H^4}{4\pi^2}e^{\pi\xi_\mathrm{eff}}z^{4+\Sigma}
\Big|c_1 W_\Sigma+c_2 M_\Sigma \Big|^2,
\label{PBB Sigma}
\\
\tilde{\mcP}_{EE}^{+(\sigma)}(z,\theta_{\bm k})&=
\frac{H^4}{4\pi^2}e^{\pi\xi_\mathrm{eff}}z^{4+\Sigma}
\left|c_1 W_\Sigma'+c_2 M_\Sigma'+\frac{\Sigma}{2z}\left(c_1 W_\Sigma+c_2 M_\Sigma\right) \right|^2,
\label{PEE Sigma}
\\
\tilde{\mcP}_{BE}^{+(\sigma)}(z,\theta_{\bm k})&=
\frac{H^4}{4\pi^2}e^{\pi\xi_\mathrm{eff}} z^{4+\Sigma}
\Big(c_1 W_\Sigma+c_2 M_\Sigma \Big)
\left(c_1 W_\Sigma'+c_2 M_\Sigma'+\frac{\Sigma}{2z}\left(c_1 W_\Sigma+c_2 M_\Sigma\right) \right)^*,
\label{PBE Sigma}
\end{align}
and $\tilde{\mcP}_{EB}^{+(\sigma)}=(\tilde{\mcP}_{BE}^{+(\sigma)})^*$.
Note that these power spectra depend on $\theta_{\bm k}$ through $\Sigma$ and $\xi_\mathrm{eff}$.
The physical energy densities of these perturbations are given by
\begin{equation}
    \rho_B=\frac{1}{4}\int_{-1}^1 \dd \cos \theta\,\int^{2\xi}_0 \frac{\dd z}{z}\,  \tilde{\mcP}_{BB}^{+(\sigma)}(z,\theta),
    \qquad
    \rho_E=\frac{1}{4}\int_{-1}^1 \dd \cos \theta\, \int^{2\xi}_0 \frac{\dd z}{z}\, \tilde{\mcP}_{EE}^{+(\sigma)}(z,\theta),
    \label{eq: rhoE and rhoB}
\end{equation}
with an UV cutoff at $-k\tau= 2\xi$, which contains all the relevant modes subject to the instability and the backreaction from the fermion production.\footnote{Precisely speaking, $\rho_{E/B}$ with an UV cutoff at $-k\tau= 2\xi$ contains both mean part and perturbation part. Here we assume that the contribution from the single mode of interest (i.e., the perturbation part) is subdominant and approximate $\tilde{E}_0/\tilde{B}_0 \sim \sqrt{2\rho_{E/B}}$ for any $k$ mode. Although this approximation breaks down when the mode becomes dominant, we do not expect substantial correction since our assumption holds for most of the evolution of the modes. We discuss this issue more in the next section.}
%
\begin{figure}[tbp]
  \includegraphics[width=80mm]{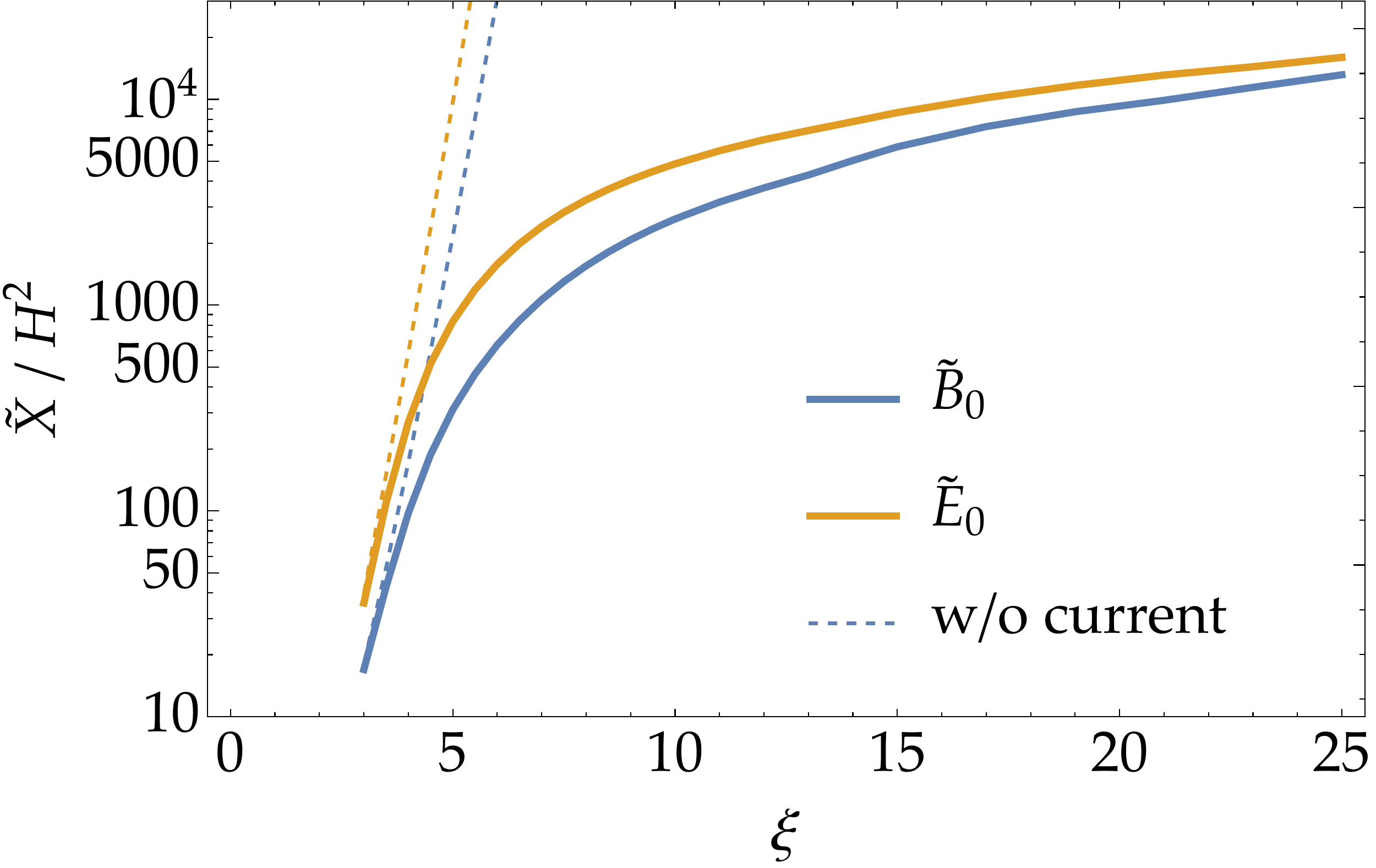}
  \hspace{8mm}
  \includegraphics[width=80mm]{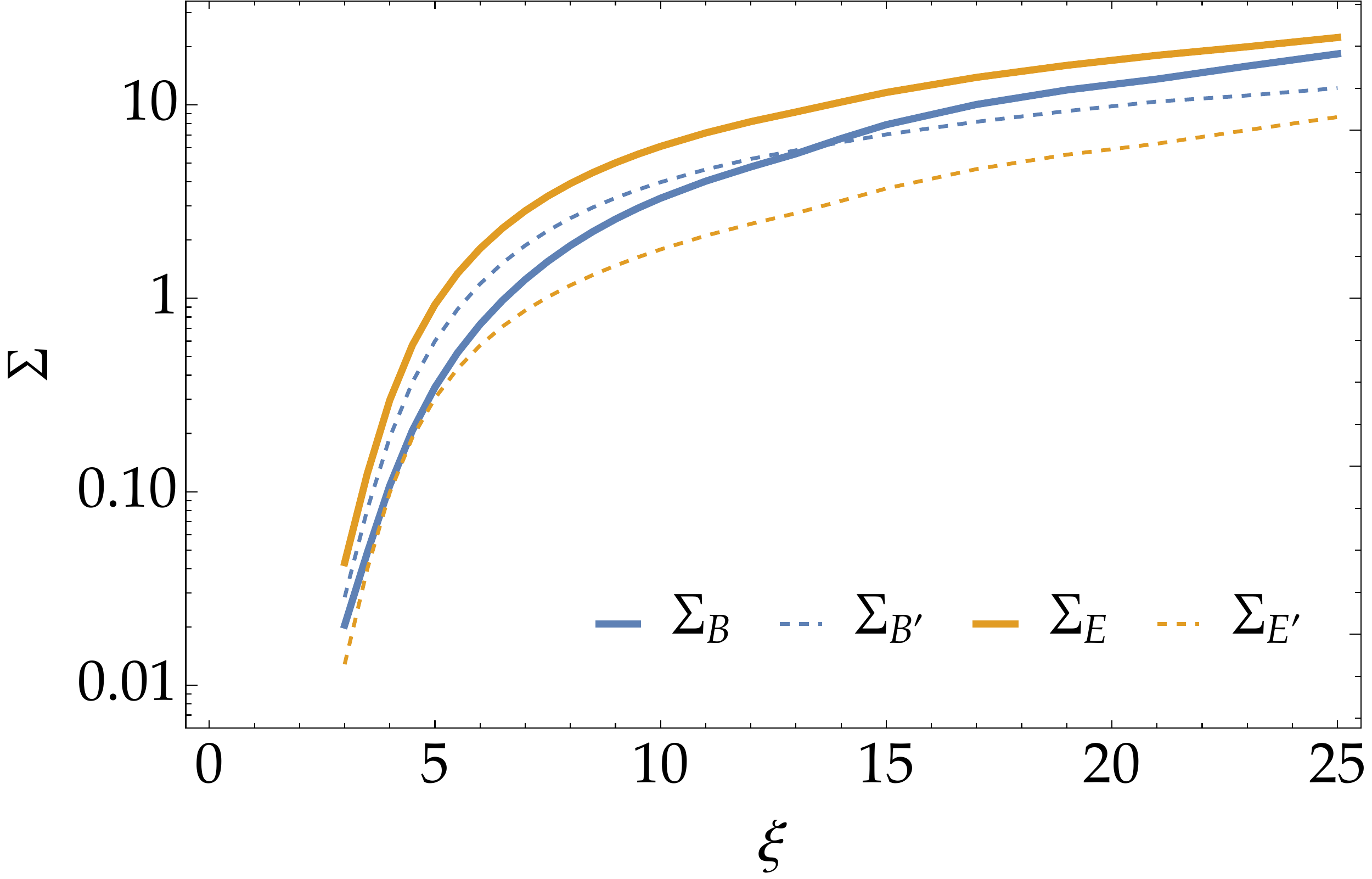}
  \caption
 {
 {\it (Left panel)}\, The self-consistent values of $\tilde{B}_0$ (blue) and $\tilde{E}_0$ (orange) in the $H^2$ unit  
 against $\xi$ obtained by numerically solving the self-consistency condition~\eqref{eq: consistency condition}. Dashed lines denote the same quantities without the current effect.
 {\it (Right panel)}\,  
 The corresponding conductivity parameters $\Sigma_B$ (blue), $\Sigma_{B^\prime}$ (blue dashed), $\Sigma_E$ (orange), and $\Sigma_{E^\prime}$ (orange dashed) defined by Eqs.~\eqref{Sigma1}--\eqref{Sigma4}.}
 \label{ConsisSigma}
\end{figure}
%

Now let us seek the self-consistent solution through the condition~\eqref{eq: consistency condition}.
We numerically find the solution as follows. First, the background $\tilde{E}_0$ and $\tilde{B}_0$ are initialized by some random values. The output $\tilde{E}_0$ and $\tilde{B}_0$ are then calculated through Eqs.~\eqref{eq: consistency condition} and \eqref{Sigma1}--\eqref{eq: rhoE and rhoB}. If the input and output values are different, the next input values are chosen between the previous input and output values. This procedure is repeated until the output values coincide with the input ones within $1\%$ errors.
The left panel of Fig.~\ref{ConsisSigma} illustrates the resultant self-consistent
background values, $\tilde{E}_0$ and $\tilde{B}_0$ while the right panel shows corresponding conductivity parameters~\eqref{Sigma1}--\eqref{Sigma4}.
We take $e=0.55$ which roughly corresponds to the coupling constant of the 
hyper $\Uone$ gauge interaction in the SM at energy scale $E\sim 10^{14}\, {\rm GeV}$ computed by two-loop renormalization group evolution (see, e.g., Ref.~\cite{Martin:1997ns}).  
Unlike the SM, however, we consider one species of Dirac fermions coupled only to the $\Uone$ gauge field for simplicity. 
One finds that the effect of the charged particles drastically suppresses the electromagnetic amplitudes and turns their exponential dependence on $\xi$ into a moderate one.
Consequently, the fermion scale $E_{\psi} \sim \sqrt{eE}$ becomes comparable to $2\xi H$. For example, $\sqrt{eE} \simeq 50H$ and $2\xi H = 20H$ with $\xi = 10$. This result a posteriori justifies our practical choice of the connection at $\gamma = 2\xi$.
Let us briefly compare our result with the previous studies. In Refs.~\cite{Domcke:2018eki, Domcke:2019mnd}, the authors estimates the maximal bound for the amplitude of the electromagnetic fields based on the stationarity of the system. 
By comparing the left panel of Fig.~6 in Ref.~\cite{Domcke:2018eki}, one can see that our self-consistent amplitude of electromagnetic fields are within this bound.\footnote{Note that our value of the gauge coupling $e = 0.55$ is different 
from the one used in Ref.~\cite{Domcke:2018eki}. This makes the maximal bound for the magnitude $\tilde{X}^2/H^4$ larger by a factor 5 in our case according to the coupling dependence in Eq.~(4.18) in Ref.~\cite{Domcke:2018eki}}
This is in contrast with Ref.~\cite{Gorbar:2021zlr} where the amplitude of electric field tends to slightly exceed the maximal bound as shown in the left panel of Fig.~1 in Ref.~\cite{Gorbar:2021zlr}.
In addition, our self-consistent amplitude of the magnetic field becomes comparable to that of the electric field when $\xi$ gets large. Such a behavior was not observed in the equilibrium estimation for the field amplitude in Refs.~\cite{Domcke:2018eki, Domcke:2019mnd}, which is based on the perturbed induced current of $e \delta \bm{J} = \sigma_B \delta \bm{B}$.

\begin{figure}[tbp]
  \includegraphics[width=80mm]{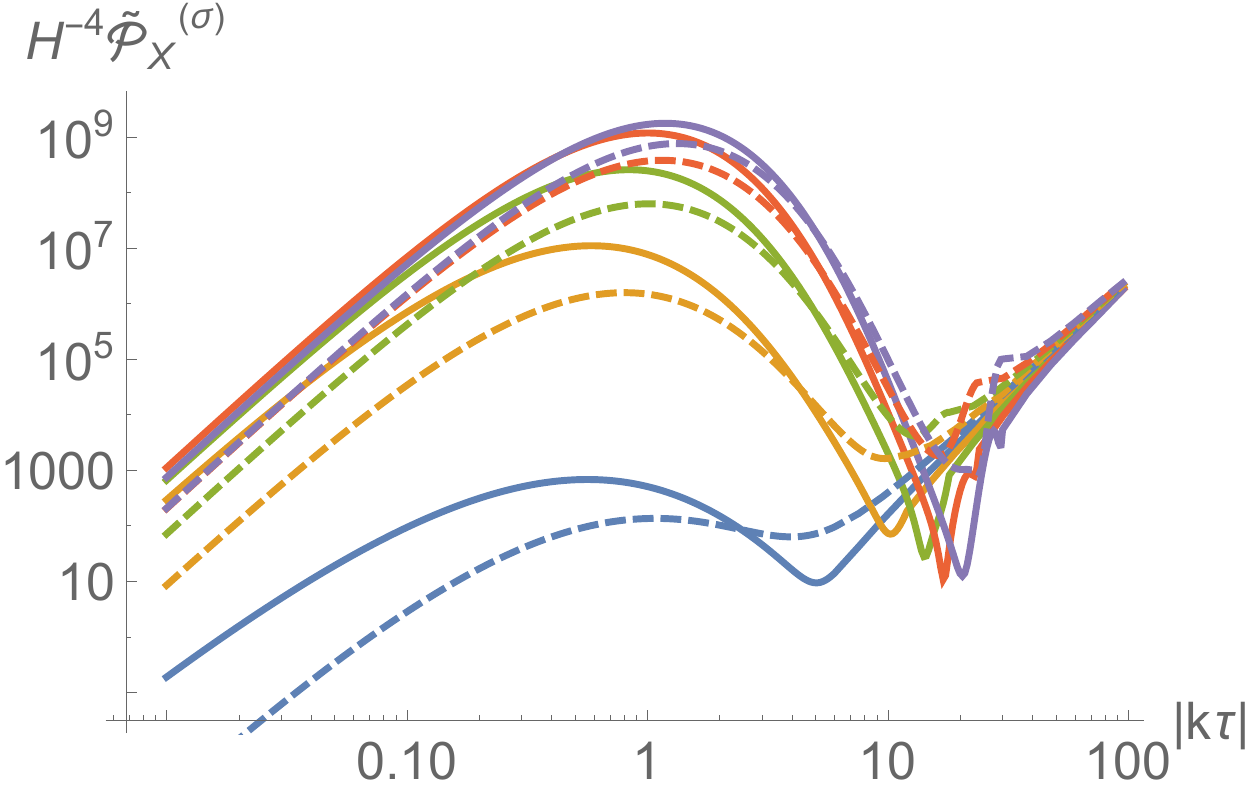}
  \hspace{8mm}
  \includegraphics[width=80mm]{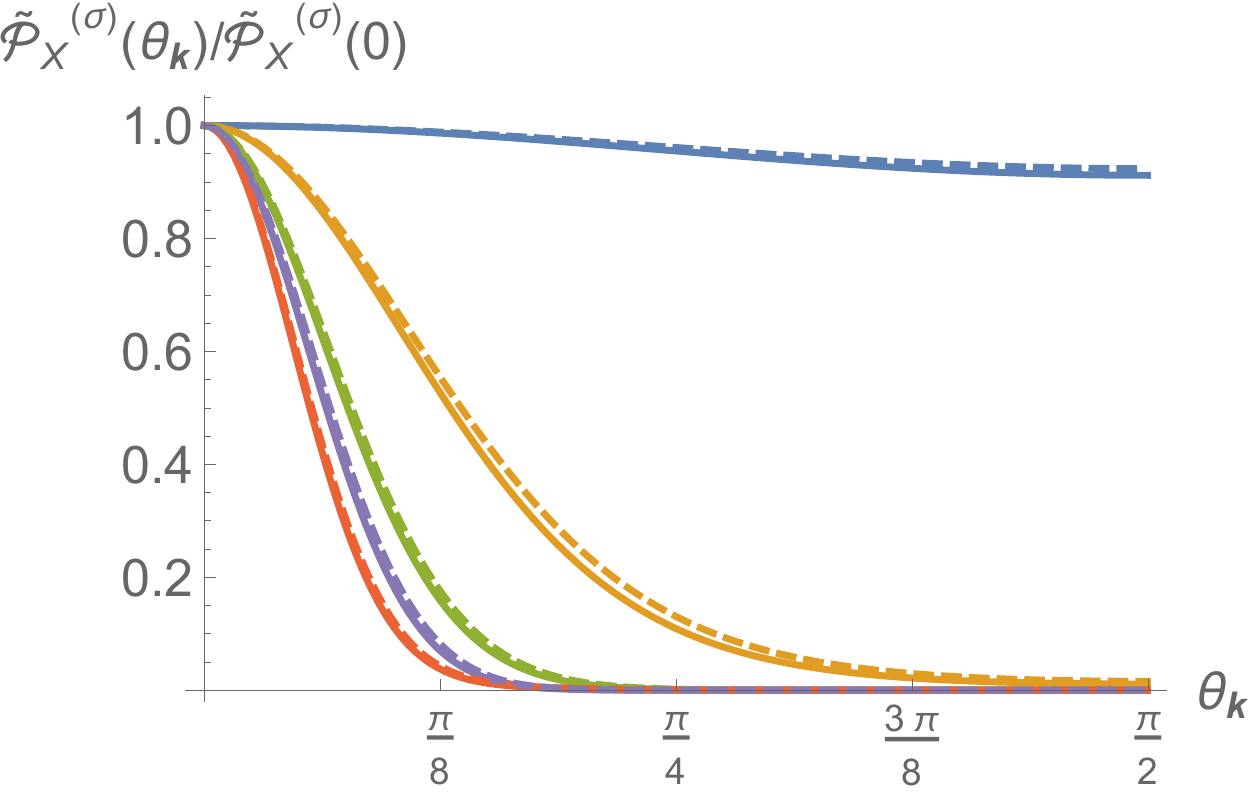}
  \caption
 {{\it (Left panel)}
 $H^{-4}\tilde{\mcP}_{EE}^{+(\sigma)}$ (solid) and $H^{-4}\tilde{\mcP}_{BB}^{+(\sigma)}$ (dashed) at the angle of maximum amplitude, $\theta_{\bm k}=0$. We adopt the consistent conductivity parameters as $\{\xi, \Sigma_E, \Sigma_E', \Sigma_B, \Sigma_B'\}=\{3, 0.043, 0.0128, 0.0204, 0.0285\}$ (blue), $\{6, 1.81, 0.571, 0.733, 1.19\}$ (orange), $\{9, 5.017, 1.48, 2.57, 3.30\}$ (green), $\{12, 8.20, 2.43, 4.78, 5.26\}$ (red) and $\{15, 11.6, 3.70, 7.90, 7.03\}$ (purple).
 {\it (Right panel)} Their angular dependence at the times of maximum amplitude. We normalize them by the values at $\theta_{\bm k}=0$. The plot scheme is the same as the left panel.}
 \label{Conductiv}
\end{figure}

In Fig.~\ref{Conductiv}, we present the electromagnetic power spectra, $\tilde{\mcP}_{EE}^{+(\sigma)}(-k\tau, \theta_{\bm k})$ and $\tilde{\mcP}_{BB}^{+(\sigma)}(-k\tau, \theta_{\bm k})$, with the self-consistent conductivity parameters. 
One observes in the left panel that the power spectra reach their peak values slightly earlier as $\xi$ increases, contrary to the case without the charged fermions where the peak scale is $|k\tau|\simeq \xi^{-1}$. This is mainly caused by the effective friction from the induced current in Eq.~\eqref{eq: A+sigma in x}. Note that these spectra still possess the properties (i)--(iv) introduced in Sec.~\ref{The U(1) gauge field without charged particles}, which validates the discussions based on them.
Moreover, the right panel shows that the angular dependence tends to be enhanced as $\xi$ increases because $\Sigma_E'$ and $\Sigma_B'$ become larger. Although the angular dependence of $\xi=15$ is slightly weaker than that of $\xi=12$, it gets stronger again for a larger $\xi$.
This result implies that under the influence of the background electromagnetic fields, new perturbation fields are most likely produced in the direction perpendicular to the background fields (cf.\ $\hat{\bm E}_0 \cdot \bm{e}^\pm (\hat{\bm k}) \propto \sin\theta_{\bm k}= 0$ for $\theta_{\bm k}=0$). It is intuitively reasonable because the induced current preventing the production of the perturbations flows parallel to the background fields.
%
%

\section{Consistency checks of the approximations}\label{consistency check}

In this section, for consistency checks of our treatment, we scrutiny four equations; (I) $\partial_\tau(a^2 e\delta \bm{J}) \propto a^4$ , (II) $\tilde{\bm E}_0\cdot \tilde{\bm B}_0=-\tilde{E}_0 \tilde{B}_0$, (III) $\tilde{\bm E}_0\cdot \tilde{\bm J}_0=\tilde{E}_0 \tilde{J}_0$,
and (IV) the 
stationarity of the gauge field energy density $\dot{\rho}_{A}=0$.
The first two equations are assumed in deriving the expression for $\delta \bm{J}$~\eqref{dJ in real space}. The latter two are expected to hold when the mean field approximation is valid.
We also present the energy distribution between the electromagnetic fields and the induced current which is transferred from the inflaton.

\subsection{Perturbed expression for the induced current}\label{pert_current}

Here we check the validity of the approximated expression for the Schwinger current~\eqref{dJ in real space}.
Let us first investigate the equation~(I), $\partial_\tau(a^2 e\delta \bm{J}) \propto a^4$. Assuming only that the background $\tilde{\bm B}_0$ and $\tilde{\bm E}_0$ are anti-parallel and static over the fermion time-scale $t_\psi$, one can obtain the perturbed version of Eq.~\eqref{Dcurrent} as
\beae{
    \partial_{\tau}(a^2e\delta \bmJ) &\simeq \frac{e^3}{2\pi^2}\left[\frac{E_0^2B_0\delta\bmE-B_0^2E_0\delta\bmB}{E_0^2+B_0^2}\coth\pqty{\frac{\pi B_0}{E_0}}\right. \\
    &\qquad\left.+\pqty{\frac{B_0^3\delta E_z-E_0^3\delta B_z}{E_0^2+B_0^2}\coth\pqty{\frac{\pi B_0}{E_0}}+(B_0\delta E_z+E_0\delta B_z)\frac{\pi B_0}{E_0}\csch^2\pqty{\frac{\pi B_0}{E_0}}}\bme_z\right].\label{diffeq_dJ}
}
This equation reproduces Eq.~\eqref{dJ in real space},
if one additionally assumes that
all $E_0$, $B_0$, $\delta E$, and $\delta B$ are proportional to $a^2$ over the Hubble time, $t_{\rm em}\sim H^{-1}$, and hence $\partial_\tau(a^2 e\delta \bm{J}) \propto a^4$.
However, as we saw in Fig.~\ref{Conductiv}, $\delta E$ and $\delta B$ experience a tachyonic growth, and thus their evolution is not simply given by $\propto a^2$. Nevertheless, we below confirm that the approximation~\eqref{dJ in real space} is justifiable by comparing with the exact solution of the differential equation~\eqref{diffeq_dJ}.

It is more convenient to express Eq.~\eqref{dJ in real space} and Eq.~\eqref{diffeq_dJ} in the momentum space for the comparison:
\begin{align}
    a^2e\delta \hat{\bmJ}_{\bm k}^{(\app)}\cdot {\bm e}^{-}(\hat{\bm k}) &\simeq \left(\Sigma_E+\Sigma_{E^{\prime}}\sin^2\theta_{\bm k}\right)\frac{\partial_{\tau}}{\tau}\hat{A}_+(\tau,{\bm k}) +\left(\Sigma_B+\Sigma_{B^{\prime}}\sin^2\theta_{\bm k}\right)\frac{k}{\tau}\hat{A}_+(\tau,{\bm k}),\label{approx}\\
    \partial_{\tau}(a^2e\delta \hat{\bmJ}_{\bm k}^{(\num)})\cdot {\bm e}^{-}(\hat{\bm k}) &\simeq 3aH\left\{\left(\Sigma_E+\Sigma_{E^{\prime}}\sin^2\theta_{\bm k}\right)\frac{\partial_{\tau}}{\tau}\hat{A}_+(\tau,{\bm k}) +\left(\Sigma_B+\Sigma_{B^{\prime}}\sin^2\theta_{\bm k}\right)\frac{k}{\tau}\hat{A}_+(\tau,{\bm k})\right\}\label{exact}.
\end{align}
We explicitly distinguish them by the superscript $(\app)$ or $(\num)$.
One can define the spectrum for the {\it physical} induced current as
\begin{align}
\tilde{\mathcal{P}}_{J} \equiv a^{-2}\mathcal{P}_{J} = \frac{k^3}{2\pi^2a^2}|\mathcal{J}_+(\tau,{\bm k})|^2,
\end{align}
where the mode function $\mathcal{J}_+(\tau,{\bm k})$ is defined as
\bae{
    e\delta \hat{\bmJ}_{\bm k}\cdot {\bm e}^{-}(\hat{\bm k}) \equiv \hat{a}_{\bm k}^{(+)}\mathcal{J}_+(\tau,{\bm k})+\hat{a}_{-{\bm k}}^{(+)\dagger}\mathcal{J}^*_+(\tau,{\bm k}).
}
Note that here we have neglected the contribution from $\mathcal{A}_-^{(\sigma)}$. In order to check the validity of the approximate formula for the induced current Eq.~\eqref{approx}, we compare the spectra analytically evaluated from Eq.~\eqref{approx} and the one numerically evaluated from Eq.~\eqref{exact} with the self-consistent solution for the gauge field $\mcA_+^{(\sigma)}$. From Eq.~\eqref{approx}, one can obtain
\bae{
    \mathcal{J}_+^{(\app)}(\tau,{\bm k}) = \frac{H^2}{\sqrt{2k}}z\left\{\Sigma\partial_z -\left(\Sigma_B+\Sigma_{B^{\prime}}\sin^2\theta_{\bm k}\right)\right\}\left(e^{\pi\xi_\mathrm{eff}/2} z^{\Sigma/2}\Big[ c_1 W_{\Sigma}(z)+c_2 M_{\Sigma}(z)\Big]\right),
}
which results in the analytical expression for the spectrum as
\bae{
    \tilde{\mathcal{P}}_{J}^{(\app)} = \frac{H^6}{4\pi^2}z^4\left|\left\{\Sigma\partial_z -\left(\Sigma_B+\Sigma_{B^{\prime}}\sin^2\theta_{\bm k}\right)\right\}\left(e^{\pi\xi_\mathrm{eff}/2} z^{\Sigma/2}\Big[ c_1 W_{\Sigma}(z)+c_2 M_{\Sigma}(z)\Big]\right)\right|^2.
    \label{Jspec_app}
}
On the other hand, one can numerically evaluate $\mathcal{J}_+^{(\num)}(\tau,{\bm k})$ from Eq.~\eqref{exact} as
\bae{
    \mathcal{J}_+^{(\num)}(\tau,{\bm k}) = \frac{3H^2}{\sqrt{2k}}z^2\int_{z}^{2\xi}\dd{z^{\prime}}\frac{1}{z'}\left\{\Sigma\partial_{z'} -\left(\Sigma_B+\Sigma_{B^{\prime}}\sin^2\theta_{\bm k}\right)\right\}\left(e^{\pi\xi_\mathrm{eff}/2} z^{\prime\Sigma/2}\Big[ c_1 W_{\Sigma}(z')+c_2 M_{\Sigma}(z')\Big]\right)\label{Jspec_num}
}
and compute the spectrum $\tilde{\mathcal{P}}_{J}^{(\num)}$. Here we take the upper limit of the integral to be $2\xi$ where we connect the consistent solution to the vacuum one. 

\begin{figure}[tbp]
  \includegraphics[width=50mm]{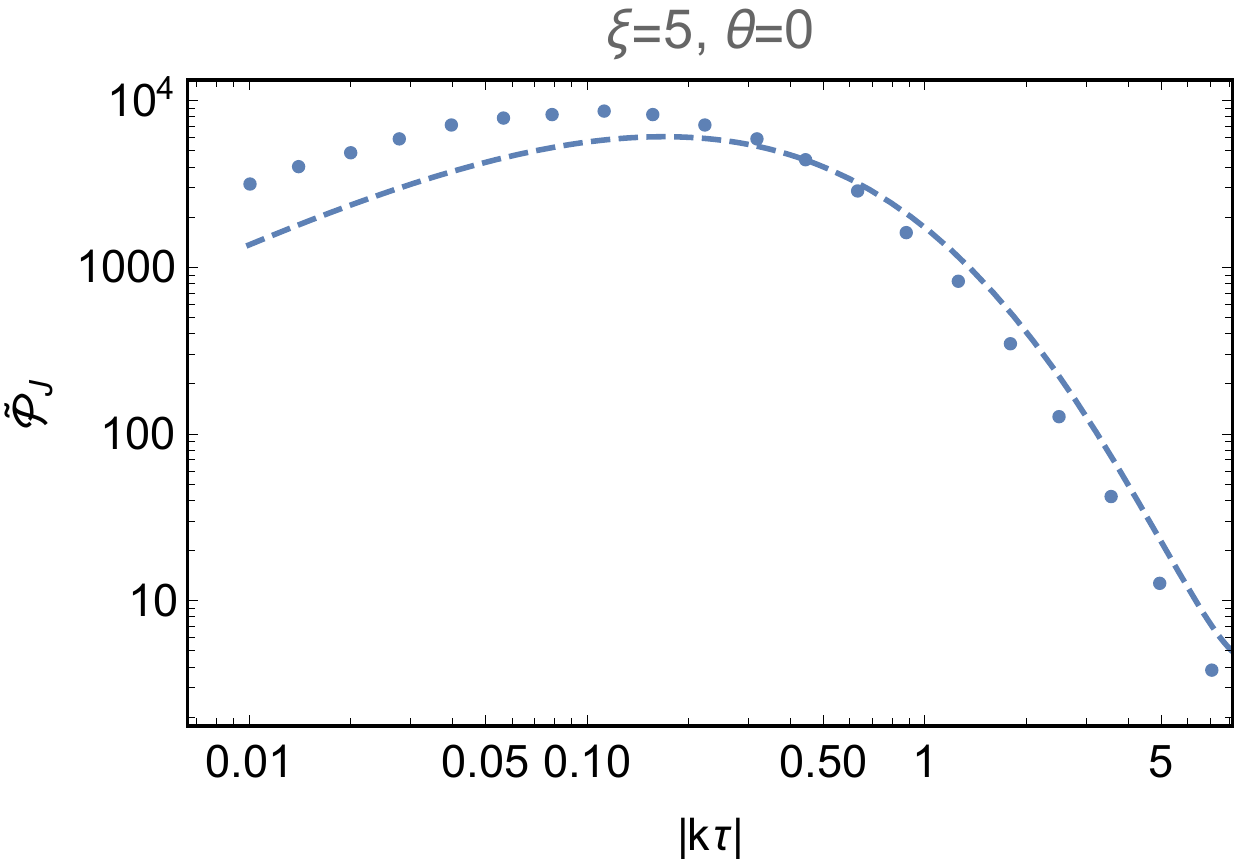}
  \hspace{3mm}
  \includegraphics[width=50mm]{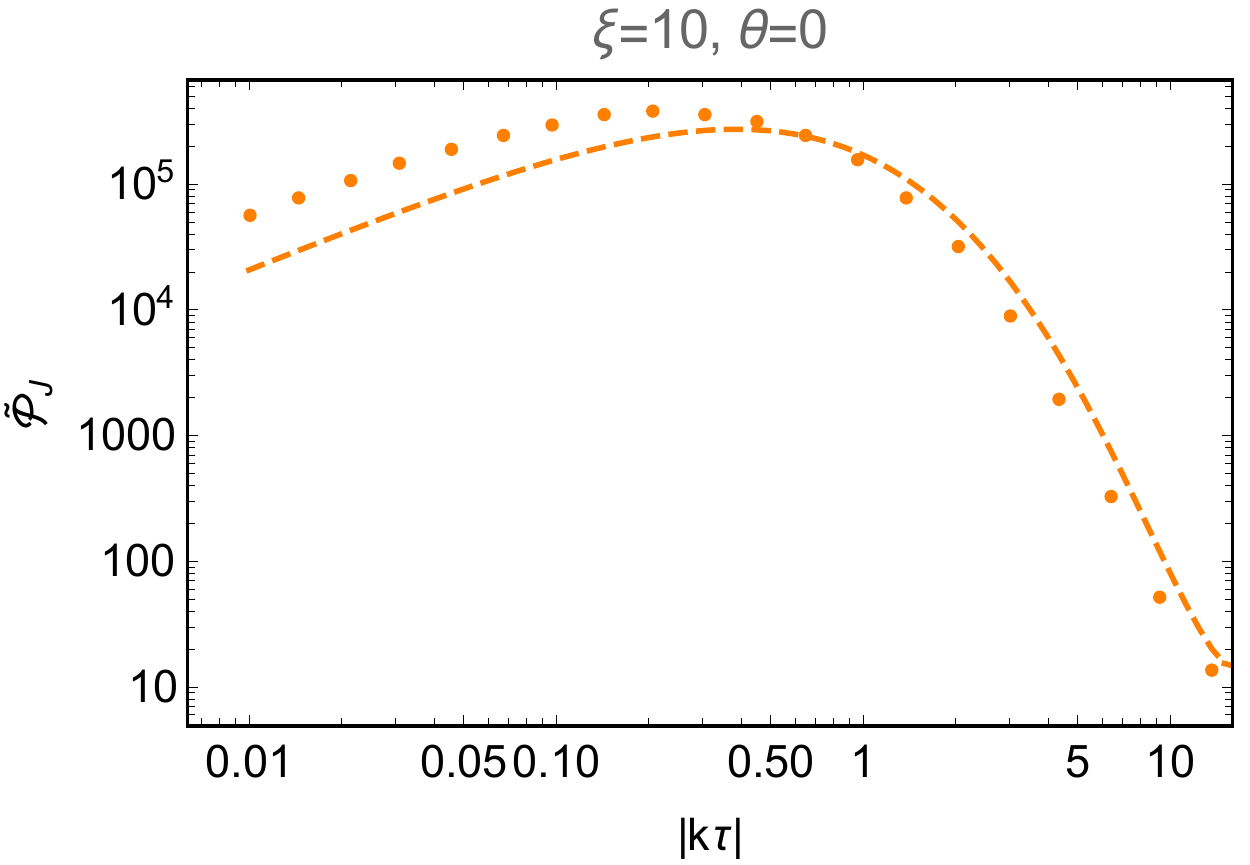}
  \hspace{3mm}
  \includegraphics[width=50mm]{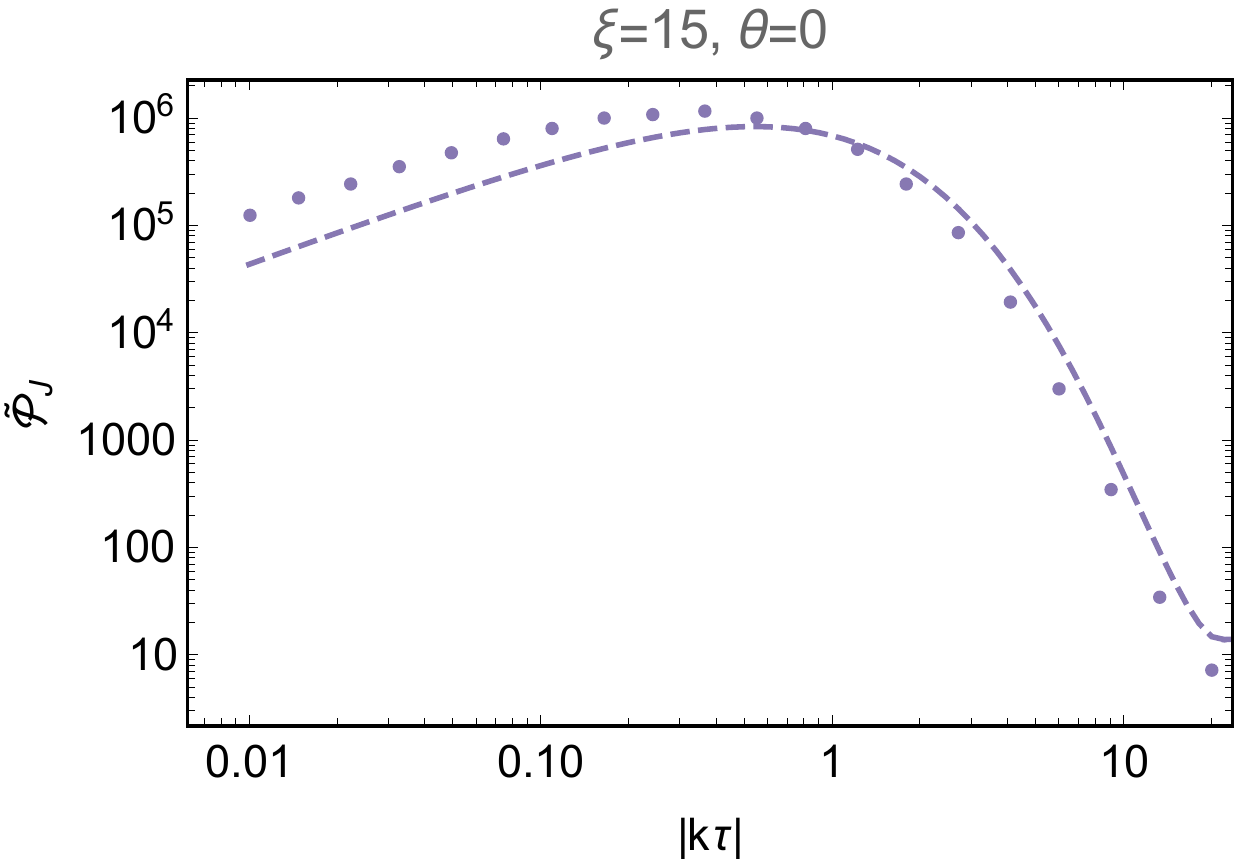}\\
  \includegraphics[width=50mm]{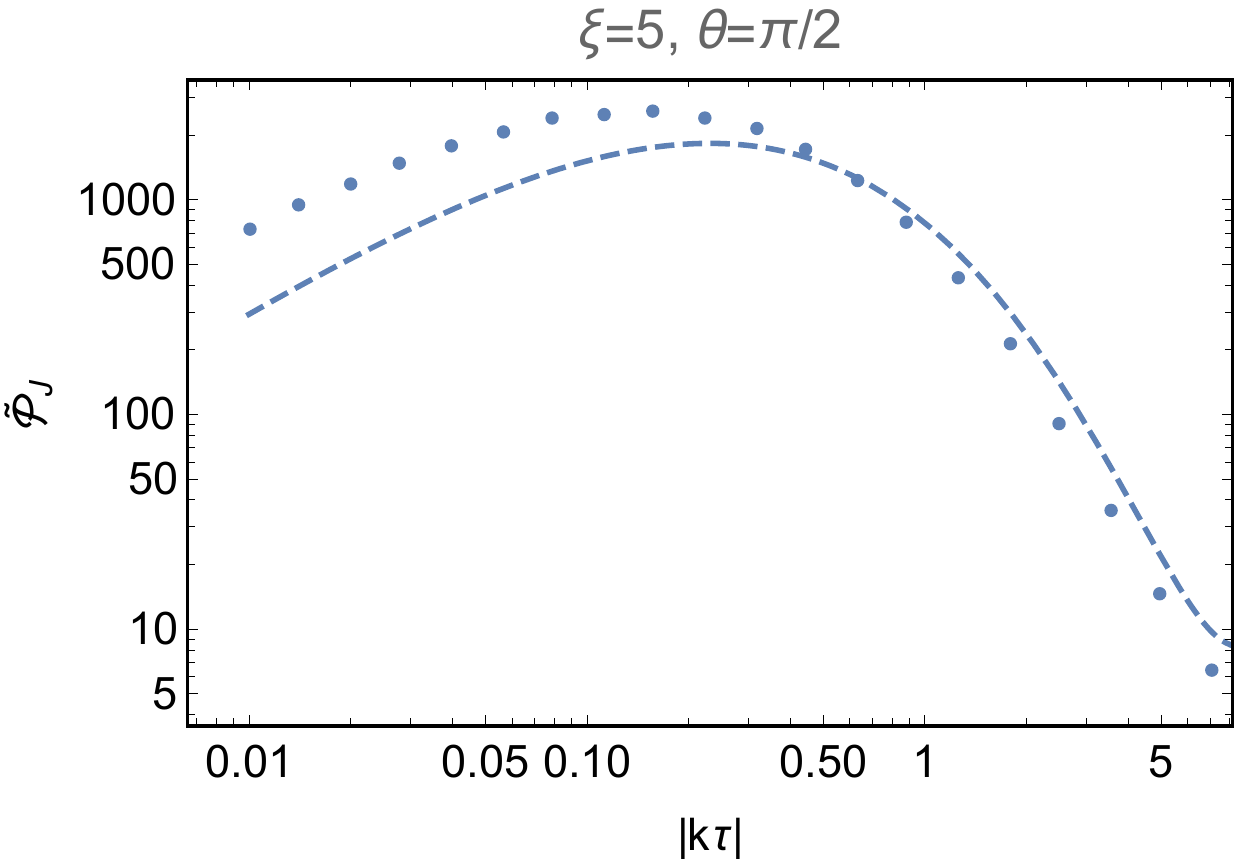}
  \hspace{3mm}
  \includegraphics[width=50mm]{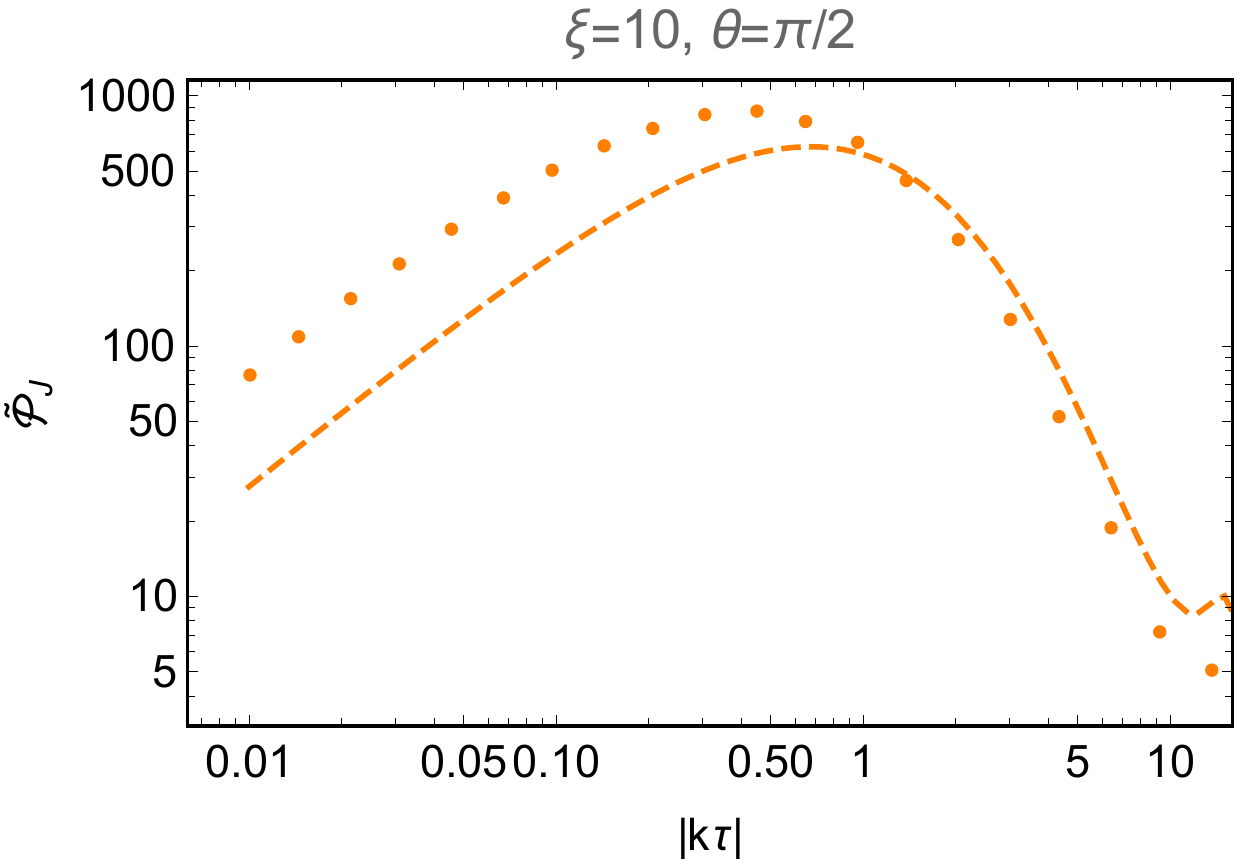}
  \hspace{3mm}
  \includegraphics[width=50mm]{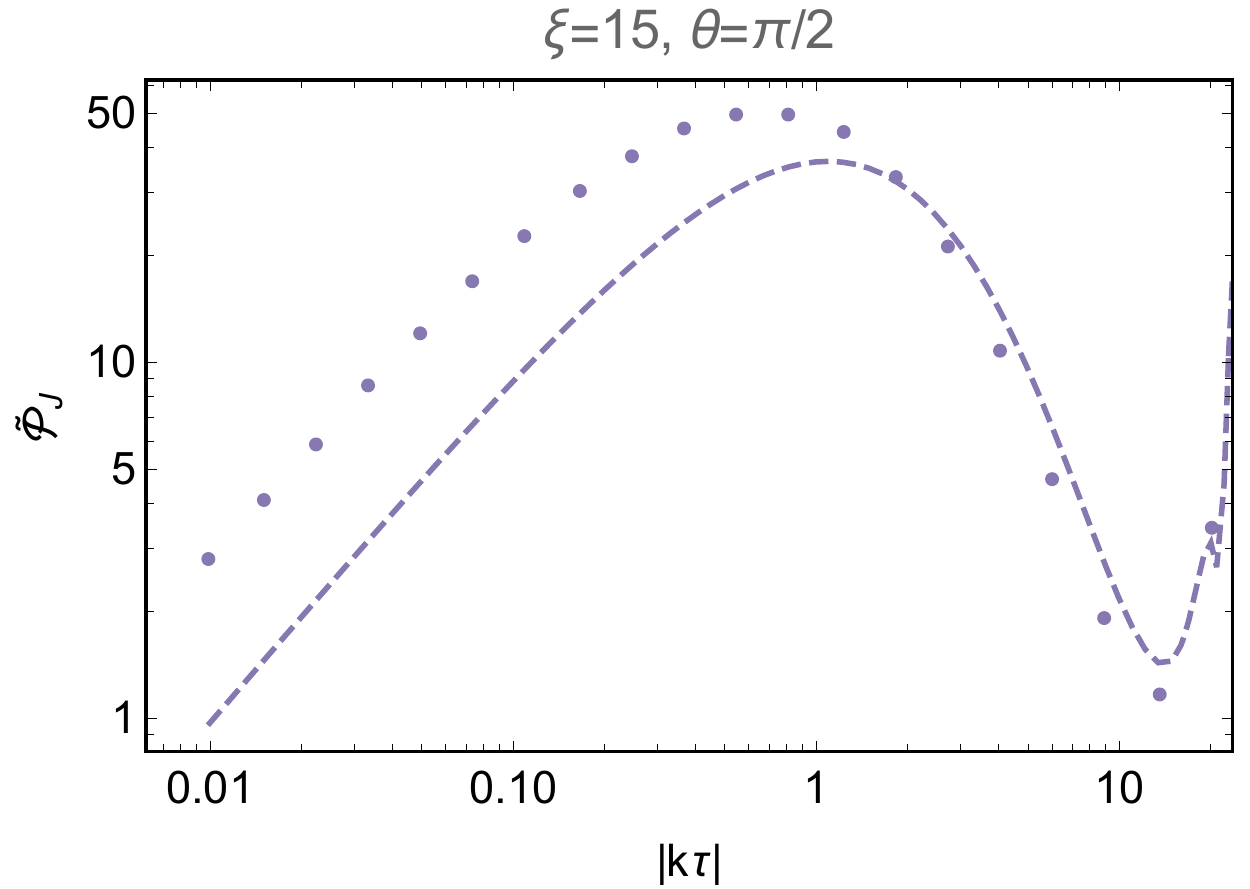}
  \caption
 {Physical current spectrum with approximated formula~\eqref{Jspec_app} (dashed line) and with numerical evaluation of Eq.~\eqref{Jspec_num} (dots) are shown. One can see that our approximate formula well reproduces the numerical values for $|k\tau|\gtrsim 1$ where the gauge field grows.
 While they differ from one another for $|k\tau|\ll 1$, the induced current does not play an important role in the super-horizon regime
 and it should not be a serious problem for our purpose.
}
 \label{Jspec}
\end{figure}

In Fig.~\ref{Jspec}, we compare $\tilde{\mathcal{P}}_{J}^{(\app)}$ and $\tilde{\mathcal{P}}_{J}^{(\num)}$ with varying $\xi$ and $\theta_{\bm k}$.
Independently of $\xi$ and $\theta_{\bm k}$, the approximated formula~\eqref{Jspec_app} 
and the numerical evaluation~\eqref{Jspec_num} show similar behaviors. In particular, they agree well for $|k\tau|\gtrsim 1$, while $\mathcal{O}(1)$ discrepancies are seen for $|k\tau|\ll 1$. We expect that these discrepancies have little impact on finding the self-consistent electromagnetic amplitudes in Sec.~\ref{self-consistent evolution}. This is because the gauge field stop evolving after the exponential growth, $\mcA_+^{(\sigma)}\simeq$ const. for $|k\tau|\ll 1$, and then the induced current does not affect its evolution. Moreover, since the physical spectra rapidly decay, the contributions from the super-horizon modes ($|k\tau|\ll 1$) to $\rho_B$ and $\rho_E$ are subdominant. On the other hand, the effect of the induced current on the evolution of the gauge field is the most important during its growing phase, $2\xi>|k\tau|\gtrsim 1$, which determines the peak amplitudes of the electromagnetic fields. Thus, Fig.~\ref{Jspec} confirms the validity of our approximation in the growing regime and justifies the use of the induced current formula~\eqref{Schwinger conductivity}.
%
%

Next, let us check the equation~(II), the anti-parallel assumption $\tilde{\bm{E}}_0\cdot\tilde{\bm{B}}_0=-\tilde{E}_0\tilde{B}_0$.
Our expressions for $\delta\bm{J}$, Eqs.~\eqref{dJ in real space} and \eqref{diffeq_dJ}, assume it, while we do not demand that the background fields obtained in the mean field approximation satisfy $\hat{\bm E}_0\cdot \hat{\bm B}_0= -1$.
In the left-panel of Fig.~\ref{rho_dot}, we present the integrated $E$-$B$ cross spectrum 
\bae{
\left\langle \tilde{\bm E}\cdot \tilde{\bm B}\right\rangle &=
    \frac{1}{4}\int_{-1}^1 \dd \cos \theta\,\int^{2\xi}_0 \frac{\dd z}{z}
    \left[\tilde{\mcP}_{EB}^{+(\sigma)}(z,\theta)+\tilde{\mcP}_{BE}^{+(\sigma)}(z,\theta)\right],
    \label{FulldotEB}
}
which is normalized by the mean field amplitude $\tilde{E}_0=\sqrt{2\rho_E}$ and $\tilde{B}_0=\sqrt{2\rho_B}$.  
Since we have identified the mean part as the summation of all the perturbation modes, we
interpret the integrated $E$-$B$ cross spectrum~\eqref{FulldotEB} as the inner product of the mean electromagnetic field $\tilde{\bm E}_0\cdot \tilde{\bm B}_0$. 
However, as seen in the left panel of Fig.~\ref{rho_dot}, the self-consistent solution is not completely anti-parallel for smaller $\xi$.
This $\xi$ dependence of the anti-parallelism of the background fields can be understood as follows. As discussed in Eq.~\eqref{Antiparalel}, only the modes for $|k\tau|\ll 2\xi$ contribute to the anti-parallel configuration.
As depicted in the left panel of Fig.~\ref{Conductiv}, the peak amplitudes of the power spectra of the field significantly grow as $\xi$ increases while the position of the peak is mostly unchanged. This means that, for large $\xi$, the mean part is dominated by the modes with $|k\tau| \sim 1 \ll 2\xi$ whose phase rotation has terminated. As a consequence, the anti-parallel assumption on the mean part holds. 
In contrast, the contributions from $|k\tau| \sim 2\xi$ cannot be neglected for smaller $\xi$ and the mean part deviates from the anti-parallel configuration.
Nevertheless, the deviation is at most about $10\%$ and hence our calculation based on $\tilde{\bm{E}}_0\cdot\tilde{\bm{B}}_0=-\tilde{E}_0\tilde{B}_0$ gives a reasonable estimate even for a small $\xi$.

\subsection{Consistency of the mean field approximation}\label{bg_consistency}

We examine the validity of the mean field approximation by investigating the equations~(III) $\tilde{\bm E}_0\cdot \tilde{\bm J}_0=\tilde{E}_0 \tilde{J}_0$ and (IV) $\dot{\rho}_A=0$ in order.
Under the mean field approximation, 
we have imposed the self-consistent condition~\eqref{eq: consistency condition} to the amplitudes $B_0$ and $E_0$. 
However, we do not impose any condition between $\bmJ_0$ and the summation of $\delta\bmJ$, which should coincide with each other if the approximation is valid.
Instead of directly comparing $|\bmJ_0|^2$ with the integration of the current spectrum $\tilde{\mathcal{P}}_J^{(\rm app)}$ in Eq.~\eqref{Jspec_app}, here we compare the inner product of the electric field and the current, which is a physical quantity relevant to the energy transfer as discussed below. 

\begin{figure}[tbp]
  \includegraphics[width=80mm]{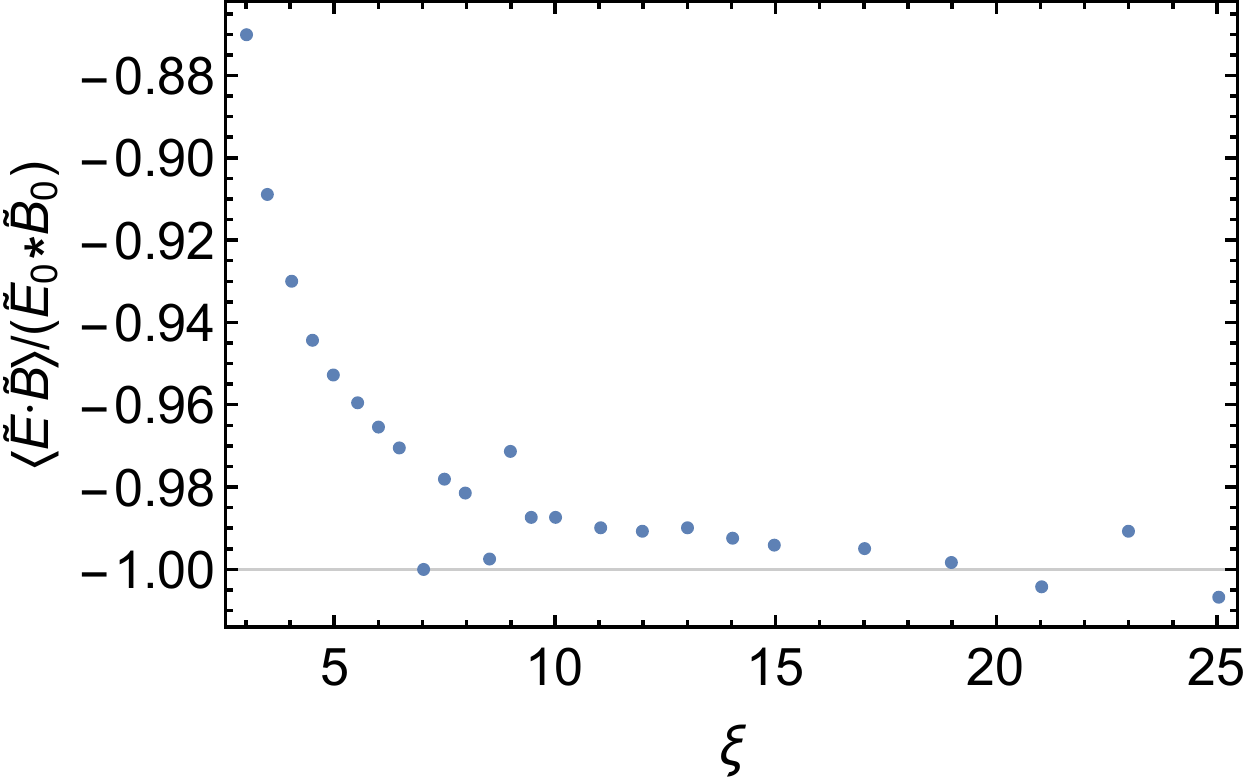}
  \hspace{8mm}
  \includegraphics[width=80mm]{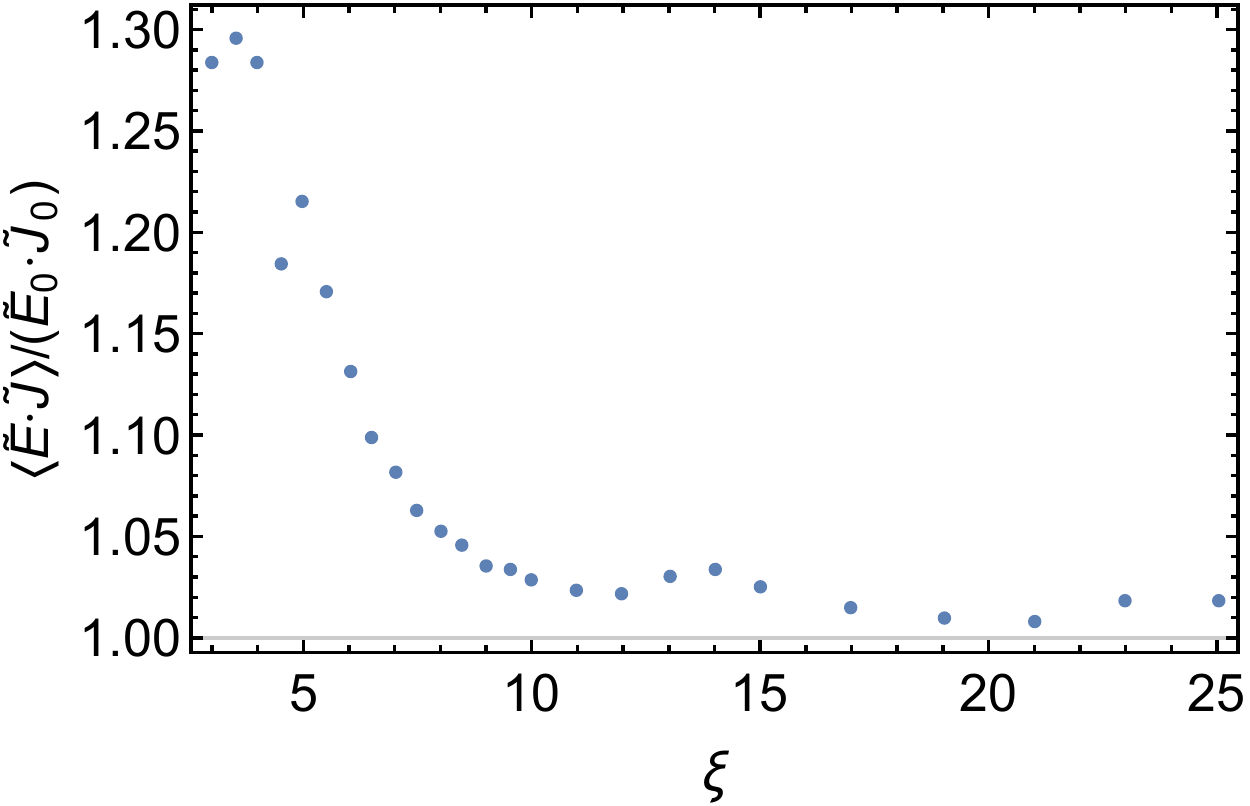}
  \caption
 { {\it (Left panel)} The degree of anti-parallelism is evaluated through the $E$-$B$ cross correlation~\eqref{FulldotEB} normalized by $\tilde{E}_0\tilde{B}_0=2\sqrt{\rho_E\rho_B}$. The anti-parallel assumption on the mean field is valid for a large $\xi$.
 {\it (Right panel)} 
 The two different expressions for the mean value of $\tilde{\bm E}\cdot\tilde{\bm{J}}$, Eqs.~\eqref{EJ_mean} and \eqref{EJ}, are compared. They agree with sufficient accuracy for a large $\xi$. However, for $\xi=\mathcal{O}(1)$, our mean field expressions have $\mathcal{O}(10\%)$ errors in the both panels.
 }
 \label{rho_dot}
\end{figure}

Using Eqs.~\eqref{dJ in real space} and \eqref{Sigma1},
one can derive 
\begin{equation}
e\tilde{\bm{E}}_0\cdot\tilde{\bm{J}}_0 = H\Sigma_E(\tilde{E}_0^2+\tilde{B}_0^2).
\label{EJ_mean}
\end{equation}
In the same way as Eq.~\eqref{FulldotEB}, we identify the following integration of the self-consistent solution with $e\tilde{\bm E}_0\cdot\tilde{\bm J}_0$:
\begin{align}
        e\left\langle \tilde{\bm E}\cdot \tilde{\bm J}\right\rangle
    &=\frac{1}{2}H \int_{-1}^1 \dd \cos \theta (\Sigma_E+\Sigma_{E^{\prime}}\sin^2\theta_{\bm k})\,\int^{2\xi}_0 \frac{\dd z}{z} \tilde{\mcP}_{EE}^{+(\sigma)}(z,\theta)
    \notag\\
    &\quad-\frac{1}{4}H \int_{-1}^1 \dd \cos \theta(\Sigma_B+\Sigma_{B^{\prime}}\sin^2\theta_{\bm k})\, \int^{2\xi}_0 \frac{\dd z}{z}
    \left[\tilde{\mcP}_{EB}^{+(\sigma)}(z,\theta)+\tilde{\mcP}_{BE}^{+(\sigma)}(z,\theta)\right].
\label{EJ}
\end{align} 
In the right-panel of Fig.~\ref{rho_dot}, we plot the ratio of these two different expressions for $e\tilde{\bm E}_0\cdot\tilde{\bm J}_0$, Eqs.~\eqref{EJ_mean} and \eqref{EJ}.
Again, they agree well for a large $\xi$. Such behavior can be understood as follows. The power spectra of the fields $\tilde{\mathcal{P}}_X^{(\sigma)}$ show the stronger angular dependence for larger $\xi$ and the dominant contributions come from $\theta_{\bm k}\simeq 0$, as depicted in Fig.~\ref{Conductiv}. 
Then, the terms with $\Sigma_{E^{\prime}/B^{\prime}}\sin^2\theta_{\bm k}$ can be ignored in Eq.~\eqref{EJ}, and the right hand side reads $H\left(\Sigma_E \tilde{E}_0^2+\Sigma_B \tilde{E}_0\tilde{B}_0\right) = H\Sigma_E \left(\tilde{E}_0^2+\tilde{B}_0^2\right)=e\tilde{\bm{E}}_0\cdot\tilde{\bm{J}}_0$. 
We note that the integration of $\tilde{\mathcal{P}}_J^{\rm (app)}$ in Eq.~\eqref{Jspec_app} also approaches to $H^2\left(\Sigma_E \tilde{E}_0^2 + 2\Sigma_E\Sigma_B\tilde{E}_0\tilde{B}_0+\Sigma_B^2\tilde{B}_0^2\right)$ $=H^2\Sigma_E^2\left(\tilde{E}_0^2+\tilde{B}_0^2\right)/\tilde{E}_0^2 = |\tilde{\bmJ}_0|^2$ for larger $\xi$, and thus reproduces the mean current amplitude as well. For a small $\xi$, on the other hand,  the discrepancy between Eqs.~\eqref{EJ_mean} and \eqref{EJ} becomes non-negligible. Although up to $30\%$ deviations are observed in the right panel of Fig.~\ref{rho_dot}, its effect on the dynamics of the entire system is limited. This is because the induced current is small in the small $\xi$ regime, as we will quantitatively see below.
%
%

Finally, let us check whether the obtained mean field is an equilibrium solution which balances the energy transfer by studying the equation~(IV), $\dot{\rho}_A=0$.
In our system, the growth of perturbations sourced by the inflaton, their dilution due to the cosmic expansion, and the backreaction from the Schwinger current  
are balanced in a complicated manner, and it is not easy to directly check their relations.
We hence confirm the consistency by
focusing on the energy transfer.
The energy density of the gauge field is given by
\begin{equation}
    \rho_A
    = \frac{1}{2}a^{-4}\left[(\partial_\tau A_i)^2 + (\partial_j A_i)^2-\partial_i A_j \partial_j A_i \right].
\end{equation}
Note that the CS coupling does not contribute to the energy density.
Taking the time derivative and the volume average, we obtain~\cite{Domcke:2018eki}
\begin{align}
\langle\dot{\rho}_A\rangle= -2H\langle\tilde{\bm E}^2+\tilde{\bm B}^2\rangle-2\xi H \langle\tilde{\bm E}\cdot \tilde{\bm B}\rangle-e \langle\tilde{\bm E}\cdot \tilde{\bm J}\rangle,
\label{rho A dot}
\end{align}
where we used the EoM for $A_i$. 
The first, second, and third terms on the right hand side in Eq.~\eqref{rho A dot} denote the dilution of the electromagnetic fields, the energy injection from the inflaton through the CS coupling, and the energy drain to the charged fermions, respectively.
When the total energy is conserved, these three terms are cancelled out. 

\begin{figure}[tbp]
  \includegraphics[width=80mm]{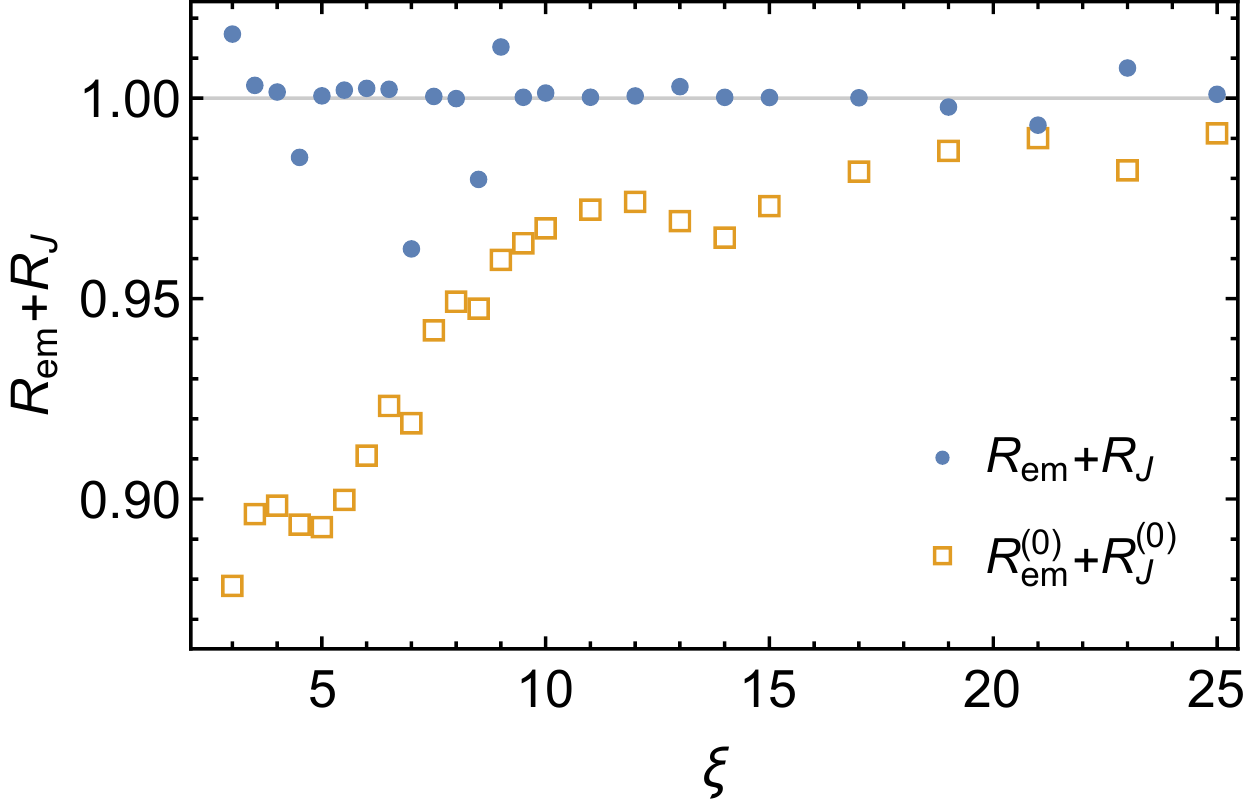}
  \hspace{8mm}
  \includegraphics[width=80mm]{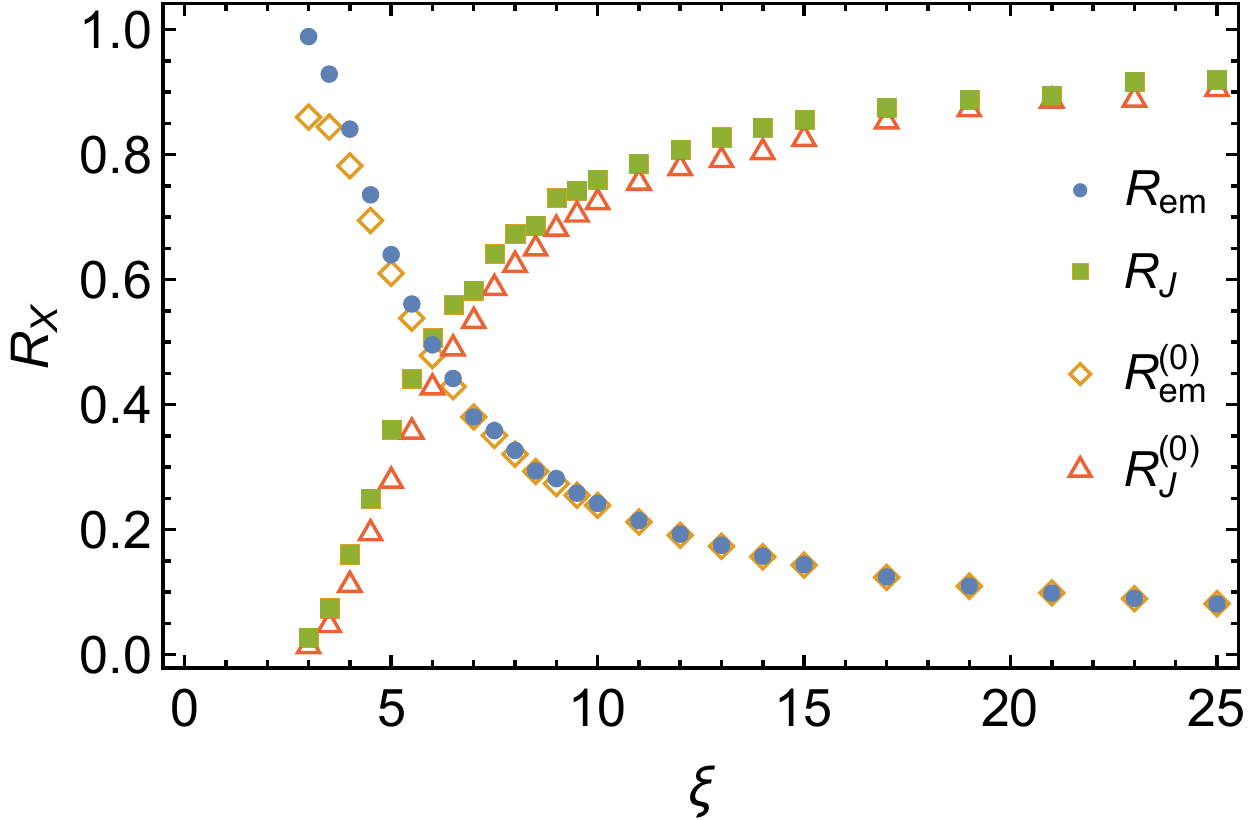}
  \caption
 {{\it (Left panel)} The blue dot represents the accuracy of the stationarity condition for full field, $R_{\rm em}+R_J=1$, given in Eq.~\eqref{Rdistribution_full}.
 Except for a few cases, the sub-percent level accuracy of the stationarity condition is achieved as expected.
 The orange square represents that for the background field, $R_{\rm em}^{(0)}+R_J^{(0)}$, which shows the violation of the stationarity condition in small $\xi$.
 {\it (Right panel)}  The energy distribution between the electromagnetic fields $R_{\rm em}$ (blue dot) and the charged fermions $R_J$ (green square). That of background field $R_{\rm em}^{(0)}$ (orange diamond) and $R_J^{(0)}$ (red triangle) is also shown.  As $\xi$ increases, more energy density transferred from the inflaton is injected to the charged fermions.
}
 \label{rho_dot_full}
\end{figure}

The stationarity of the gauge field energy density, $\langle\dot{\rho}_A\rangle=0$, can be recast as
\begin{equation}
    R_{\rm em}+R_J=1,
    \qquad R_{\rm em}\equiv\frac{\langle\tilde{\bm E}^2
    +\tilde{\bm B}^2\rangle}{\xi |\langle\tilde{\bm E}\cdot \tilde{\bm B}\rangle|},
    \qquad
    R_J\equiv \frac{e\left\langle \tilde{\bm E}\cdot \tilde{\bm J}\right\rangle}{2\xi H |\langle\tilde{\bm E}\cdot \tilde{\bm B}\rangle|},
    \label{Rdistribution_full}
\end{equation}
where $R_{\rm em}$ and $R_J$ indicate the
ratio of the energy which the electromagnetic fields
and the fermion gain from the inflaton, respectively. 
In the left-panel of Fig.~\ref{rho_dot_full}, we present them evaluated with self-consistent solution.
Since it is a solution to the perturbed EoM,
our self-consistent solution satisfies Eq.~\eqref{Rdistribution_full} within a few \% accuracy as expected (recall that we allow $1\%$ errors in our numerical procedure to find the consistency solution).
%
%

Next, we evaluate $R_{\rm em}$ and $R_J$
with our static background $\tilde{\bm E}_0$ and $\tilde{\bm B}_0$. 
The stationarity condition of the mean part are written as
\begin{equation}
    R_{\rm em}^{(0)}+R_J^{(0)}=1,
    \qquad R_{\rm em}^{(0)}\equiv
    \frac{\tilde{\bm{E}}_0^2+\tilde{\bm{B}}_0^2}{\xi\abs{\tilde{\bm{E}}_0\cdot\tilde{\bm{B}}_0}}\simeq\frac{\tilde{E}_0^2+\tilde{B}_0^2}{\xi \tilde{E}_0\tilde{B}_0},
    \qquad
    R_J^{(0)}\equiv 
    \frac{e\tilde{\bm{E}}_0\cdot\tilde{\bm{J}}_0}{2\xi H\abs{\tilde{\bm{E}}_0\cdot\tilde{\bm{B}}_0}}\simeq\frac{\Sigma_E(\tilde{E}_0^2+\tilde{B}_0^2)}{2\xi\tilde{E}_0\tilde{B}_0},
    \label{Rdistribution}
\end{equation}
where we used $|\tilde{\bm E_0}\cdot \tilde{\bm B_0}| \simeq \tilde{E}_0\tilde{B}_0$
and Eq.~\eqref{EJ_mean}, namely the equations~(II) and (III), which are studied in Fig.~\ref{rho_dot}. 
The accuracy of the stationarity condition, $\langle\dot{\rho}_A^{(0)}\rangle=0$, is shown in the left panel of Fig.~\ref{rho_dot_full}. One can see that the sum of $R_{\rm em}^{(0)}$ and $R_J^{(0)}$ becomes closer to unity as $\xi$ becomes larger, which again indicates that our mean field approximation works well with a larger value of $\xi$. 
For a small $\xi$, however, $R_{\rm em}^{(0)}+R_J^{(0)}$ deviates from unity by at most about $10\%$.
This error comes from the violations of the equations~(II) and (III).
In the smallest $\xi$ regime, the $10\%$ overestimation in $\tilde{\bm E}_0\cdot \tilde{\bm B}_0$ dominates the error and the effect of the $30\%$ underestimation in $\tilde{\bm{E}}_0\cdot\tilde{\bm{J}}_0$ is invisible, because almost all energy is transferred to the electromagnetic fields.

In the right panel of Fig.~\ref{rho_dot_full}, we individually plot $R_{\rm em}$, $R_J$, and their background counterparts to present the distribution of the total energy injected from the inflaton. 
$R_J^{(0)}$ receives both the errors from the violations of the equations~(II) and (III), while $R_{\rm em}^{(0)}$ does only from the equation~(II). However, since the former becomes subdominant $R^{(0)}_J\ll R_{\rm em}^{(0)}$ for $\xi\lesssim 5$,  its deviation from $R_J$ does not significantly worsen the stationarity condition for the mean field. 
Interestingly enough, $R_J$ overwhelms $R_{\rm em}$ for $\xi\gtrsim 10$, and the charged fermions gain a dominant part of the total energy injected by the inflaton. Such fermion domination is also observed in Ref.~\cite{Domcke:2018eki}.

Since all of the four equations~(I)--(IV) are validated, especially with a sufficiently large value of $\xi$, we conclude that our treatment with the approximated expression for $\delta \bm J$ and the mean field approximation can be applied to the system. 
The equation~(I), which approximates
the time dependence of the induced current, holds independently of $\xi$ in the relevant regime where
the gauge field perturbation grows.
The others~(II)--(IV) exhibit $\mathcal{O}(10\%)$ errors for a small value of $\xi$.
Although the violation of the equation~(III), which estimates the mean current, becomes the largest for a small $\xi$, its error does not significantly propagate to the electromagnetic fields or the stationarity of the gauge field energy density.
The assumption of the anti-parallel configuration of the mean electromagnetic fields is also violated at most about $10\%$.
In order to give a more reliable prediction for smaller values of $\xi$, we may need to remove the anti-parallel assumption in all the calculations, 
which might further complexify the problem.

\section{Conclusion}\label{conclusion}

In this paper, we studied a coupled system where the axionic inflaton generates the electromagnetic fields via the CS coupling, and simultaneously the charged Dirac fermions are produced as well as accelerated by the electromagnetic fields. This model is well motivated as an inflationary model and is naturally linked to an interesting phenomenology such as baryogenesis and magnetogenesis. All the gauge symmetries in SM have associated charged particles and realistic studies on the related phenomenology need to include their effects. Hence, it is crucially important to take into account the charged particles in the analysis.
However, except for a few recent attempts~\cite{Domcke:2018eki,Domcke:2019mnd,Domcke:2019qmm,Gorbar:2021rlt,Gorbar:2021zlr}, most of the previous works neglected the backreaction from the charged particles, and hence this complicated system has not been investigated thoroughly.

We developed a procedure to obtain an equilibrium solution for the electromagnetic fields under the effects of the charged particles and the inflaton with a constant velocity in a self-consistent manner. We pointed out that there is a scale separation between the electromagnetic fields and the fermions produced by the Schwinger effect, which enables us to integrate out the fermions. Then, the induced current is described as a function of the electromagnetic fields, but it makes the EoM highly non-linear. We introduced the two approximations, constant physical electromagnetic fields and the mean field approximation, to find a linearized EoM for the perturbed gauge field, which has an analytic solution. By numerically solving the self-consistency equations, we obtained the electromagnetic spectra and the conductivity parameters and found that the current effect drastically suppressed the electromagnetic amplitudes.

We also carefully examined the validity of our approximations. For larger $\xi\equiv \dot{\phi}/(2fH)$, our treatment was validated with sufficient accuracy. When $\xi=\mathcal{O}(1)$, the approximated equations are subjected to $\mathcal{O}(10\%)$ errors. In this small $\xi$ regime, however, the energy transfer to the induced current is subdominant and the uncertainty arising by the fermions does not significantly propagate to the estimate of the electromagnetic fields. The main source of the error originates in the anti-parallel assumption of the mean electromagnetic fields and hence developing a generalized formalism without this assumption would be a fascinating future work. 
It is also interesting to compare our treatment with existing literatures Refs.~\cite{Domcke:2018eki,Domcke:2019mnd} and Refs.~\cite{Gorbar:2021rlt,Gorbar:2021zlr}.
The former study also tried to find an equilibrium configuration of the gauge field but assuming that the induced current for the perturbed mode only has the magnetic conductivity.
The latter studies introduced the spacial gradient expansion of the electromagnetic fields to treat the dynamics but assuming that the induced current only has the electric conductivity. As commented in footnote~\ref{footnote}, the prescription of the vacuum fluctuation is also different. 
In contrast, our approach focuses on finding an equilibrium solution and involves both the electric and magnetic conductivities for the perturbed equation of motion.
The comparison between these approaches would be beneficial for a better understanding of the system.

One can extend our procedure in multiple directions. First, we expect that the energy transfer from the inflaton to the gauge-fermion sector should be most efficient at the end of inflation where $\xi$ reaches its maximum value. However, the slow-roll approximation and hence our $\xi=\text{const.}$ assumption are violated there, and the stationarity condition for the gauge field is also expected to be broken. One needs to restore the inflaton from the external energy source into a dynamical field and simultaneously solve it.
Second, the $\theta_{\bm k}$ dependence of the electromagnetic spectra may imply that the mean field constantly changes its direction, which is not incorporated in our procedure. The stochastic formalism~\cite{Starobinsky:1986fx,Fujita:2017lfu,Talebian:2019opf,Talebian:2022jkb} is known to be capable of tracking the time evolution of mean fields, to which perturbations are continually added, and may be useful to accommodate the rotational behavior. Finally, we considered only one species of Dirac fermions charged under the U(1) gauge symmetry for simplicity.  The SM contains more U(1) charged particles as well as non-Abelian gauge sectors, which may also be coupled to the axionic inflaton. Adding these ingredients to our procedure would be interesting.
We leave them for future work.

\section*{Acknowledgments}
We would like to thank useful discussions with 
Kohei Kamada, Misao Sasaki, and Jun'ichi Yokoyama.
This work was supported in part by the Japan Society for the Promotion of Science (JSPS) KAKENHI, Grant Number
JP18K13537, JP20H05854 (T.F.), 
JP20J21866 (J.K.), 
JP19K14707, and JP21K13918 (Y.T.).
J.K.\ was supported by research program of the Leading Graduate Course for Frontiers of Mathematical Sciences and Physics (FMSP).
K.M.\ was supported by MEXT Leading Initiative for Excellent Young Researchers Grant Number JPMXS0320200430.

\appendix
\section{Thermalization of the fermions}
\label{Thermalization}
As discussed in Ref.~\cite{Domcke:2018eki}, one should pay careful attention to the scattering of the fermions after the Schwinger production during inflation. If the produced particles get thermalized due to the scattering, the property of the plasma is just characterized by the temperature $T$ and chiral chemical potential $\mu_5$.
This results in a expression for the current as $\tilde\bmJ = \sigma \tilde\bmE + \mu_5 \tilde\bmB$ for $\tilde E,\tilde B \ll T^2$ where the \textit{thermal} electric conductivity is $\sigma \sim T / \alpha$, which is different from the one \eqref{Dcurrent} we used.
To verify our expression of the induced current~\eqref{Dcurrent}, here we qualitatively discuss that scattering might not be frequent enough to prevent the acceleration by following the discussion in Ref.~\cite{Domcke:2018eki} with our self-consistent electromagnetic fields.

Practically, it is difficult to follow the dynamical process of thermalization.
Instead, let us adopt an extreme assumption that the produced fermions were thermalized within one Hubble time $1/H$, and see whether our estimation could be modified.
The would-be temperature under this assumption is given as
\bae{
T_\text{wb} \sim \left(\frac{30e \tilde{E} \tilde{J}}{\pi^2 g_* H} \right)^{1/4}
\sim 0.1 \times \qty( e \tilde B )^\frac{1}{4} \qty( \frac{e \tilde E}{H} )^\frac{1}{2},
}
where we used Eq.~\eqref{Schwinger conductivity} and $g_*\simeq 10^2$.
The first nontrivial check is whether the fermions can remain thermalized, \textit{once they somehow get thermalized}.
By comparing the typical scattering rate in the thermal plasma, $\alpha^2 T_\text{wb}$, and the Hubble parameter $H$, we find that the fermions remain thermalized for
\bae{
\label{thermalization_consistent}
1 \ll 0.1\, \alpha^2 \qty( \frac{e \tilde B}{H^2} )^\frac{1}{4} \qty( \frac{e \tilde E}{H^2} )^\frac{1}{2}
\longrightarrow
\frac{\tilde E}{H^2} \gg 10^6 \times \qty( \frac{0.55}{e} )^\frac{19}{3}.
}
Here we used $\tilde E \sim \tilde B$, which holds for the parameters of our interest as shown in Fig.~\ref{ConsisSigma}.

Another question is \textit{how the fermions would get thermalized}; in other words, what is the bottleneck process for the fermions to be thermalized.
If the scatterings are negligible, the typical momentum of fermions after $1/H$ is $e \tilde E / H$ owing to the acceleration by the electric field, which is much larger than the would-be temperature $T_\text{wb}$.
This efficient acceleration is implicitly assumed in the derivation of the Schwinger current given in Eq.~\eqref{Dcurrent}, and hence we need to clarify the regime of validity.
The interaction rate of such high-energy particles is suppressed by the Landau--Pomeranchuk--Migdal effect as $\Gamma_\text{LPM} \sim \alpha^2 T_\text{wb} \sqrt{T_\text{wb} / p}$ for $p \gg T_\text{wb}$ with $p$ being a momentum~\cite{Gyulassy:1993hr,Arnold:2001ba,Arnold:2002ja,Kurkela:2011ti,Harigaya:2013vwa}.
Here we quote the result in the presence of non-Abelian gauge interactions, having in mind a realistic situation based on the SM, where almost all fermions except for the right-handed leptons are charged under non-Abelian gauge fields.\footnote{
  In non-Abelian gauge field theories, the LPM suppressed rate for a high-energy particle with energy $E$ to emit a gauge field with energy $\omega$ is $\Gamma_\text{LPM} \sim \alpha^2 T \sqrt{ T / \omega }$.
  On the other hand, the same rate in Abelian gauge field theories is $\Gamma_\text{LPM}^\text{U$(1)$} \sim \alpha^2 T \sqrt{\omega T / E^2 }$. This difference originates from the fact that the non-Abelian gauge fields are charged under themselves.
  One can see that the LPM suppressed rate in Abelian gauge field theories is smaller than that in non-Abelian gauge field theories (because of $E > \omega$).
}
If this process is efficient $\Gamma_\text{LPM} \gg H$, a typical momentum after acceleration would be $e \tilde E / \Gamma_\text{LPM}$ instead of $e \tilde E / H$.
Inserting $p \sim e \tilde E / \Gamma_\text{LPM}$ back into the inequality $\Gamma_\text{LPM} (p) \gg H$, we obtain the following self-consistency condition for thermalization~\cite{Domcke:2018eki}
\bae{
\label{thermalization_lpm}
1 \ll 10^{-3}\, \alpha^4 \qty( \frac{e \tilde B}{H^2} )^\frac{3}{4}
\qty( \frac{e \tilde E}{H^2} )^\frac{1}{2}
\longrightarrow
\frac{\tilde E}{H^2} \gg 5\times10^7 \times \qty( \frac{0.55}{e} )^\frac{37}{5}.
}

In the parameters of our interest (see e.g., Fig.~\ref{ConsisSigma}), neither of these conditions, \eqref{thermalization_lpm} nor \eqref{thermalization_consistent}, is satisfied.
Thus, the acceleration is much more efficient than scatterings~\eqref{thermalization_lpm}.
Furthermore, even if the produced fermions were thermalized somehow, scatterings could not maintain thermal equilibrium~\eqref{thermalization_consistent}.
Based on these estimations, we adopt Eq.~\eqref{Dcurrent}, which is derived by neglecting scatterings, as a reasonable approximation for the induced current.

\section{Lorentz boost}
\label{Lorentz boost}

As we only know the expression of the Schwinger current for anti-parallel electromagnetic fields, let us consider the Lorentz boost to make general electromagnetic fields anti-parallel.
Note that hereafter we basically focus on the linear response in $\delta E/E_0\sim\delta B/B_0\sim\epsilon$ on the anti-parallel background $\bmE_0$ and $\bmB_0$.

Suppose the constant and homogeneous (physical) electromagnetic fields $\bmE=\bmE_0+\delta\bmE$ and $\bmB=\bmB_0+\delta\bmB$ on the anti-parallel background $\bmE_0$ and $\bmB_0$.
Without loss of generality, $\bmE$ and $\bmB$ can be assumed in the $xz$-plane as $\bmE=(E_x,0,E_z)^T$ and $\bmB=(B_x,0,B_z)^T$ (note that the background $\bmE_0$ and $\bmB_0$ are not necessarily in the $xz$-plane).
We can further assume that $\bmE$ and $\bmB$ are almost anti-parallel along the $z$-direction as $E_z>0$ and $B_z<0$, and also the $x$-axis can be chosen so that $E_x>0$ and $B_x>0$. They can be expressed with angles $\phi_E$ and $\phi_B$ as
\bae{
    \bmE=\pmqty{
        E_x \\
        0 \\
        E_z
    }=\pmqty{
        E\sin\phi_E \\
        0 \\
        E\cos\phi_E
    } \qc
    \bmB=\pmqty{
        B_x \\
        0 \\
        B_z
    }=\pmqty{
        B\sin\phi_B \\
        0 \\
        -B\cos\phi_B
    }.
}
The configuration is schematically illustrated in Fig.~\ref{fig: Lorentz}.
Note that the angles $\phi_E$ and $\phi_B$ can be expected as small as $\calO(\epsilon)$ because one could take $\phi_E=\phi_B=0$ without $\delta\bmE$ and $\delta\bmB$, which is justified later.

\bfe{width=0.7\hsize}{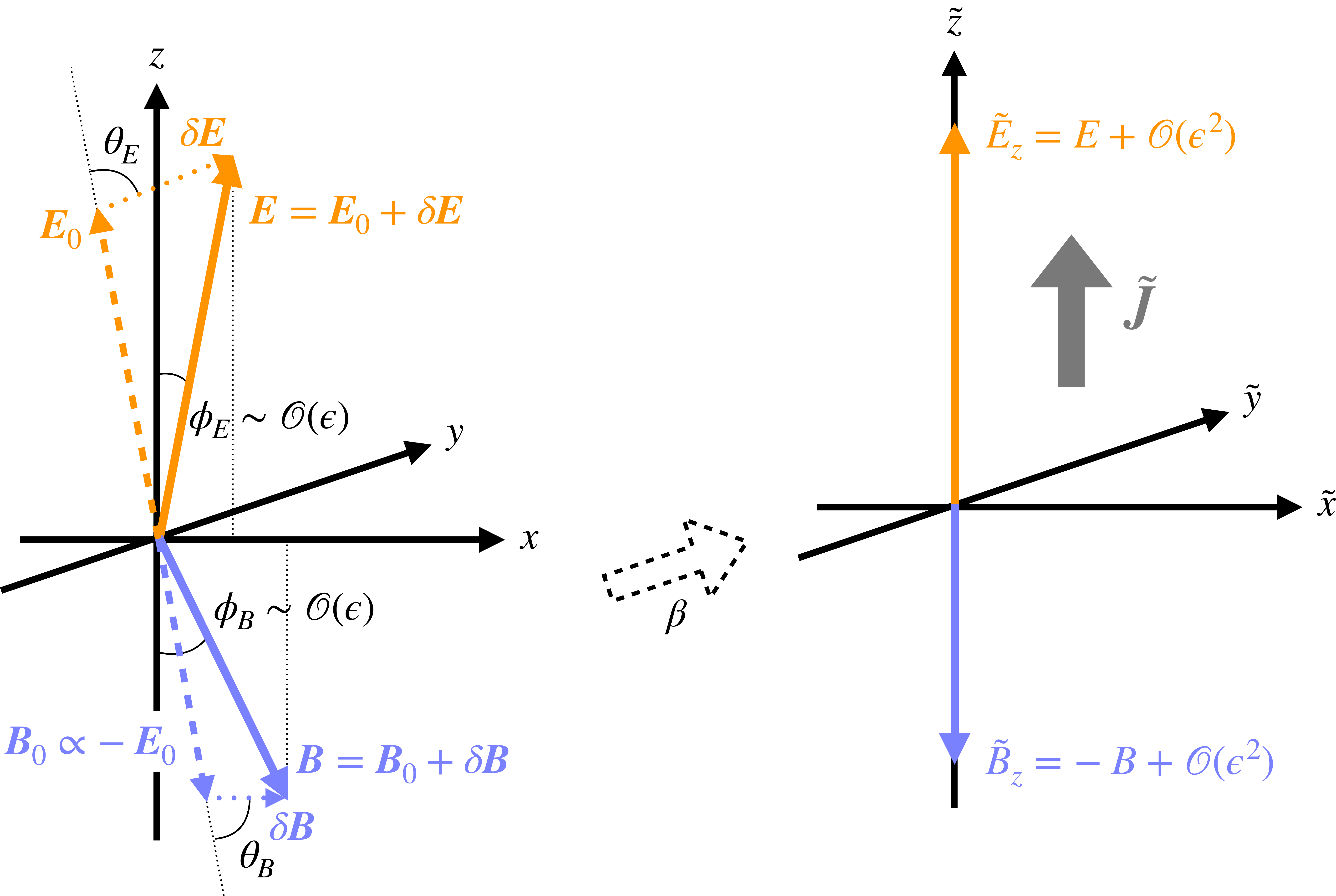}{The schematic image of the configuration of the electromagnetic fields and the Schwinger current. Electromagnetic fields in the $xz$-plane can be parallelized by a Lorentz boost in the $y$-direction. Up to linear order in $\epsilon\sim\delta E/E_0\sim\delta B/B_0$, the amplitudes of the electromagnetic fields do not change. The Schwinger current is also found to be along the $z$-axis both in the $x^\mu$ and $\tilde{x}^\mu$ coordinates with the same strength to the linear order. Note that the $z$-direction is defined to be $\hat{E}_0$ in the main body. In this definition, the current direction is found as Eq.~\eqref{eq: Jhat}.}{fig: Lorentz}

Regarding the spacetime, we can fix the scale factor as, e.g., $a=1$ in the conformal flat metric $\dd{s^2}=a^2(\tau)(\dd{\tau^2}-\dd{\bmx^2})$ for a while as the dynamics of the Schwinger current is much faster than the time scale of the universe's expansion.
Then we consider the Lorentz boost in $y$-direction:
\bae{	
	\bce{
		\dps
		\tilde{\tau}=\gamma(\tau-\beta y), \\
		\dps
		\tilde{x}=x, \\
		\dps
		\tilde{y}=\gamma(y-\beta\tau), \\
		\dps
		\tilde{z}=z,
	}
	\quad\Leftrightarrow\quad
	\bce{
		\dps
		\tau=\gamma(\tilde{\tau}+\beta\tilde{y}), \\
		\dps
		x=\tilde{x}, \\
		\dps
		y=\gamma(\tilde{y}+\beta\tilde{\tau}), \\
		\dps
		z=\tilde{z},
	}
}
where $\gamma=1/\sqrt{1-\beta^2}$. The electromagnetic fields transform as
\beae{
    &\bce{
        \dps
        \tilde{E}_x=\gamma(E_x+\beta B_z), \\
        \dps
        \tilde{E}_y=E_y, \\
        \dps
        \tilde{E}_z=\gamma(E_z-\beta B_x),
    } \quad
    \bce{
        \dps
        \tilde{B}_x=\gamma(B_x-\beta E_z), \\
        \dps
        \tilde{B}_y=B_y, \\
        \dps
        \tilde{B}_z=\gamma(B_z+\beta E_x),
    } \\
    \Leftrightarrow\quad&\bce{
        \dps
        E_x=\gamma(\tilde{E}_x-\beta\tilde{B}_z), \\
        \dps
        E_y=\tilde{E}_y, \\
        \dps
        E_z=\gamma(\tilde{E}_z+\beta\tilde{B}_x),
    } \quad
    \bce{
        \dps
        B_x=\gamma(\tilde{B}_x+\beta\tilde{B}_z), \\
        \dps
        B_y=\tilde{B}_y, \\
        \dps
        B_z=\gamma(\tilde{B}_z-\beta\tilde{E}_x),
    }
}
which can be checked as the transformation of the field strength tensor
\bae{
    F_{\mu\nu}=\pmqty{
        0 & E_x & E_y & E_z \\
        -E_x & 0 & -B_z & B_y \\
        -E_y & B_z & 0 & -B_x \\
        -E_z & -B_y & B_x & 0
    }.
}
We then seek the Lorentz boost with which the transformed electromagnetic fields become anti-parallel in the $z$-direction as $\tilde{\bmE}=(0,0,\tilde{E}_z)^T$ and $\tilde{\bmB}=(0,0,\tilde{B}_z)^T$.

The values of $\tilde{E}_z$ and $\tilde{B}_z$ themselves are easily obtained, making use of the Lorentz invariance of $F_{\mu\nu}F^{\mu\nu}=-2(\bmE^2-\bmB^2)$ and $\epsilon^{\mu\nu\rho\sigma}F_{\mu\nu}F_{\rho\sigma}=-8\bmE\cdot\bmB$, as
\bae{
    \tilde{E}_z^2=\frac{E^2-B^2+\sqrt{(E^2-B^2)^2+4E^2B^2\cos^2\phi_{EB}}}{2} \qc
    \tilde{B}_z^2=\frac{2E^2B^2\cos^2\phi_{EB}}{E^2-B^2+\sqrt{(E^2-B^2)^2+4E^2B^2\cos^2\phi_{EB}}},
}
where $\phi_{EB}=\pi-\phi_E-\phi_B$ is the angle between $\bmE$ and $\bmB$. However, as one can expect $\phi_E\sim\phi_B\sim\calO(\epsilon)$, $\cos\phi_{EB}$ can be approximated by unity to the linear order because $\cos\phi_{EB}=1+\calO\pqty{(\pi-\phi_{EB})^2}=1+\calO(\epsilon^2)$.
One then finds that $\tilde{E}_z$ and $\abs{\tilde{B}_z}$ are equivalent to the original lengths $E$ and $B$ to the linear order:
\bae{
    \tilde{E}_z\simeq E\simeq E_0+\delta E\cos\theta_E \qc
    \tilde{B}_z\simeq-B\simeq-B_0-\delta B\cos\theta_B.
}
Here we also show the expansion of the original lengths $E$ and $B$ with use of the angles $\theta_E$ and $\theta_B$ between $\bmE_0$ and $\delta\bmE$ and between $\bmB_0$ and $\delta\bmB$. Note that $\theta_E$ and $\theta_B$ are not necessarily small in contrast to $\phi_E$ and $\phi_B$.

In the $\tilde{x}^\mu$ coordinate, the electromagnetic fields are anti-parallel in the $\tilde{z}$-direction with the amplitudes $E$ and $B$, and therefore the evolution of the Schwinger current in this coordinate is expected as~\cite{Domcke:2018eki} (see also Eq.~\eqref{Dcurrent})
\bae{
    -\partial_{\tilde{\tau}}e\tilde{J}^\mu=\bce{
        \dps
        \frac{e^3BE}{2\pi^2}\coth\pqty{\frac{\pi B}{E}}, & \mu=\tilde{z}, \\
        0 & \text{otherwise}.
    }
}
Its inverse boost reads
\bae{
    -(\gamma\partial_\tau+\gamma\beta\partial_y)eJ^\mu=\bce{
        \dps
        \frac{e^3BE}{2\pi^2}\coth\pqty{\frac{\pi B}{E}}, & \mu=z, \\
        0 & \text{otherwise}.
    }
}
We then suppose the expansion of the current as $J^\mu=J_0^\mu+\delta J^\mu$ with the background current $eJ_{0i} = -a^2eJ_0^i=\frac{e^3B_0E_{0i}}{6\pi^2a^3H}\coth\pqty{\frac{\pi B_0}{E_0}}$.
Noting that the boost parameter $\beta$ is also expected to be $\calO(\epsilon)$ and $J_0^\mu$ is spatially homogeneous by definition, one finds that the current equation reduces to
\bae{
    -\partial_\tau eJ^\mu\simeq\bce{
        \dps
        \frac{e^3BE}{2\pi^2}\coth\pqty{\frac{\pi B}{E}}, & \mu=z, \\
        0, & \text{otherwise},
    }
}
to the linear order.
Restoring the cosmic expansion as $-J^i\to-a^4J^i=a^2J_i$ (see Eq.~\eqref{int lagrangian}), and assuming the physically constant electromagnetic fields as $E,B\propto a^2$, this current equation can be solved as
\bae{
    a^2eJ_\mu\simeq\bce{
        \dps
        \frac{e^3BE}{6\pi^2aH}\coth\pqty{\frac{\pi B}{E}}, & \mu=z, \\
        0, & \text{otherwise}.
    }
}
The Schwinger current thus flows in the $z$-direction even in the original coordinate to the linear order. 

We have to then specify the $z$-direction in the $\bmE\bmB$-plane, i.e., the angles $\phi_E$ and $\phi_B$.
They are obtained by the conditions $\tilde{E}_x=\tilde{B}_x=0$. 
Recalling $\beta\sim\calO(\epsilon)$, one finds
\bae{
    \bce{
        \dps
        0=\tilde{E}_x=\gamma(E_x+\beta B_z)\simeq E_0\phi_E-\beta B_0, \\
        \dps
        0=\tilde{B}_x=\gamma(B_x-\beta E_z)\simeq B_0\phi_B-\beta E_0,
    }
}
to the leading order.
Therefore the boost parameter $\beta$ is given by
\bae{
    \beta=\frac{E_0}{B_0}\phi_E=\frac{B_0}{E_0}\phi_B.
}
In other words, two angles $\phi_E$ and $\phi_B$ are determined as
\bae{\label{eq: phiE phiB}
    \phi_E=\frac{B_0^2}{E_0^2+B_0^2}(\pi-\phi_{EB}) \qc
    \phi_B=\frac{E_0^2}{E_0^2+B_0^2}(\pi-\phi_{EB}),
}
once $\phi_{EB}$, the angle between $\bmE$ and $\bmB$, is fixed via $\pi-\phi_{EB}\simeq\sin\phi_{EB}=\abs{\bmE\times\bmB}/EB$.

For the convenience in the $A_i$'s EoM, let us redefine the $z$-direction to the $\bmE_0$-direction as $\bmE_0=(0,0,E_0)$ and $\bmB_0=(0,0,-B_0)$, and then find the current direction $\hat{\bmJ}$ in this coordinate.
As $\hat{\bmJ}$ is in the $\bmE\bmB$-plane, it can be expressed as
\bae{
    \hat{\bmJ}=a\hat{\bmE}+b\hat{\bmB},
}
with some coefficients $a$ and $b$.
Here we assume $a>0$ and $b<0$ because $\hat{\bmJ}$ is almost parallel to $\hat{\bmE}$.
The angles $\phi_E$ and $\phi_B$~\eqref{eq: phiE phiB} between $\hat{\bmJ}$ and $\hat{\bmE}$ and between $-\hat{\bmJ}$ and $\hat{\bmB}$ are related to the coefficients $a$ and $b$ via
\beae{
    &\phi_E\simeq\sin\phi_E=\abs{\hat{\bmJ}\times\hat{\bmE}}=-b\abs{\hat{\bmB}\times\hat{\bmE}}=-b\sin\phi_{EB}\simeq-b(\pi-\phi_{EB}), \\
    &\phi_B\simeq\sin\phi_B=\abs{\hat{\bmJ}\times\hat{\bmB}}=a\abs{\hat{\bmE}\times\hat{\bmB}}=a\sin\phi_{EB}\simeq a(\pi-\phi_{EB}).
}
Comparing it with Eq.~\eqref{eq: phiE phiB}, one finds
\bae{
    a=\frac{E_0^2}{E_0^2+B_0^2} \qc
    b=-\frac{B_0^2}{E_0^2+B_0^2}.
}
Therefore, making use of 
\bae{
    \hat{\bmE}\simeq\pqty{1-\frac{\delta E_z}{E_0}}\bme_z+\frac{\delta\bmE}{E_0} \qc
    \hat{\bmB}\simeq-\pqty{1+\frac{\delta B_z}{B_0}}\bme_z+\frac{\delta\bmB}{B_0},
}
the current direction is revealed as
\bae{\label{eq: Jhat}
    \hat{\bmJ}=\bqty{1-\frac{E_0\delta E_z-B_0\delta B_z}{E_0^2+B_0^2}}\bme_z+\frac{E_0\delta\bmE-B_0\delta\bmB}{E_0^2+B_0^2}.
}

\section{Derivation of the perturbed EoM for the gauge field}
\label{Derivation of the perturbed EoM for the gauge field}

Substituting Eq.~\eqref{dJ in real space} into the perturbed version of Eq.~\eqref{Ai EoM w/ full current},
we have
\beae{
&\partial_\tau^2\, \delta A_i-\partial_j^2\, \delta A_i+\frac{2\xi}{\tau} \epsilon_{ijl}\partial_j \delta A_l
\\
&=\frac{e^3}{6\pi^2aH}\left[\pqty{\frac{B_0^3\delta E_z-E_0^3\delta B_z}{E_0^2+B_0^2}\coth\pqty{\frac{\pi B_0}{E_0}}+(B_0\delta E_z+E_0\delta B_z)\frac{\pi B_0}{E_0}\csch^2\pqty{\frac{\pi B_0}{E_0}}}\bme_z \right. \\
    &\qquad\left.+\frac{E_0^2B_0\delta\bmE-B_0^2E_0\delta\bmB}{E_0^2+B_0^2}\coth\pqty{\frac{\pi B_0}{E_0}}\right].
}
Considering $\delta \bm{E} =- \partial_\tau \delta \bm A$ and $\delta \bm B=\bm{\nabla}\times \delta \bm A$, 
this is a linear equation with respect to $\delta \bm A$.
Thus, to find the EoM in Fourier space, we make the following replacements:
\begin{align}
    \delta A_i(\tau, \bm x) \to \sum_{\lambda=\pm} e_i^{(\lambda)}(\hat{\bm k})A_\lambda(\tau,\bm k),
    \quad
    \delta E_i(\tau, \bm x) \to -\sum_{\lambda=\pm} e_i^{(\lambda)}(\hat{\bm k})\partial_\tau A_\lambda(\tau,\bm k),
    \quad
    \delta B_i(\tau, \bm x) \to \sum_{\lambda=\pm} e_i^{(\lambda)}(\hat{\bm k}) \lambda k A_\lambda(\tau,\bm k).
\end{align}
We obtain
\beae{
    &\sum_{\lambda=\pm} \left[\partial_\tau^2 +k^2+2\lambda k\frac{\xi}{\tau}\right]{\bm e}^{(\lambda)}A_\lambda(\tau,\bm{k})
    \\
    &=\frac{e^3B_0}{6\pi^2a^2H^2}\pqty{\frac{B_0^2}{E_0^2+B_0^2}\coth\pqty{\frac{\pi B_0}{E_0}}+\frac{\pi B_0}{E_0}\csch^2\pqty{\frac{\pi B_0}{E_0}}}\pqty{-\frac{\sin\theta_k}{\sqrt{2}}}\pqty{\sum_{\lambda=\pm}\frac{\partial_\tau}{\tau} A_\lambda(\tau,\bm{k})}\bme_z \\
    &+\frac{e^3E_0}{6\pi^2a^2H^2}\pqty{\frac{E_0^2}{E_0^2+B_0^2}\coth\pqty{\frac{\pi B_0}{E_0}}-\frac{\pi B_0}{E_0}\csch^2\pqty{\frac{\pi B_0}{E_0}}}\pqty{-\frac{\sin\theta_k}{\sqrt{2}}}\pqty{\sum_{\lambda=\pm}\lambda\frac{k}{\tau}A_{\lambda}(\tau,\bm{k})}\bme_z\\
    &+\frac{e^3B_0}{6\pi^2a^2H^2}\left(\frac{E_0^2}{E_0^2+B_0^2}\coth\pqty{\frac{\pi B_0}{E_0}}\right)\sum_{\lambda=\pm}\frac{\partial_{\tau}}{\tau}A_{\lambda}(\tau,\bm{k})\bme^{(\lambda)}(\hat{\bm{k}})
    \\
    &+\frac{e^3E_0}{6\pi^2a^2H^2}\left(\frac{B_0^2}{E_0^2+B_0^2}\coth\pqty{\frac{\pi B_0}{E_0}}\right)\sum_{\lambda=\pm}\lambda\frac{k}{\tau}A_{\lambda}(\tau,\bm{k})\bme^{(\lambda)}(\hat{\bm{k}}).
}
Note that we take $\hat{\bm E}_{0}\cdot {\bm e}^\pm(\hat{\bm k})=\bme_z\cdot {\bm e}^\pm(\hat{\bm k})=-\sin\theta/\sqrt{2}$ as discussed in the footnote~\ref{footnote: polarizaion vector}. 
To extract each polarization mode, we multiply this equation by $e_i^\mp$ 
and use $\bm{e}^+\cdot\bm{e}^+=\bm{e}^-\cdot\bm{e}^-=0$ and $\bm{e}^+\cdot \bm{e}^-=1$.
Then the above equation is split into 
\beae{
    &\left[\partial_\tau^2 +k^2+2k\frac{\xi}{\tau}\right]A_+
    \\
    &=\frac{e^3B_0}{6\pi^2a^2H^2}\left(\frac{E_0^2}{E_0^2+B_0^2}\coth\pqty{\frac{\pi B_0}{E_0}}\right)\frac{\partial_{\tau}}{\tau}A_{+}
    \\
    &+\frac{e^3E_0}{6\pi^2a^2H^2}\left(\frac{B_0^2}{E_0^2+B_0^2}\coth\pqty{\frac{\pi B_0}{E_0}}\right)\frac{k}{\tau}A_{+}
    \\
    &+\frac{e^3B_0}{12\pi^2a^2H^2}\pqty{\frac{B_0^2}{E_0^2+B_0^2}\coth\pqty{\frac{\pi B_0}{E_0}}+\frac{\pi B_0}{E_0}\csch^2\pqty{\frac{\pi B_0}{E_0}}}\sin^2\theta_k\pqty{\sum_{\lambda=\pm}\frac{\partial_\tau}{\tau} A_{\lambda}}\\
    &+\frac{e^3E_0}{12\pi^2a^2H^2}\pqty{\frac{E_0^2}{E_0^2+B_0^2}\coth\pqty{\frac{\pi B_0}{E_0}}-\frac{\pi B_0}{E_0}\csch^2\pqty{\frac{\pi B_0}{E_0}}}\sin^2\theta_k\pqty{\sum_{\lambda=\pm}\lambda\frac{k}{\tau}A_{\lambda}},
    \label{A+EoM fullRHS}
}
and
\beae{
    &\left[\partial_\tau^2 +k^2-2k\frac{\xi}{\tau}\right]A_-
    \\
    &=\frac{e^3B_0}{6\pi^2a^2H^2}\left(\frac{E_0^2}{E_0^2+B_0^2}\coth\pqty{\frac{\pi B_0}{E_0}}\right)\frac{\partial_{\tau}}{\tau}A_{-}
    \\
    &-\frac{e^3E_0}{6\pi^2a^2H^2}\left(\frac{B_0^2}{E_0^2+B_0^2}\coth\pqty{\frac{\pi B_0}{E_0}}\right)\frac{k}{\tau}A_{-}
    \\
    &+\frac{e^3B_0}{12\pi^2a^2H^2}\pqty{\frac{B_0^2}{E_0^2+B_0^2}\coth\pqty{\frac{\pi B_0}{E_0}}+\frac{\pi B_0}{E_0}\csch^2\pqty{\frac{\pi B_0}{E_0}}}\sin^2\theta_k\pqty{\sum_{\lambda=\pm}\frac{\partial_\tau}{\tau} A_{\lambda}}\\
    &+\frac{e^3E_0}{12\pi^2a^2H^2}\pqty{\frac{E_0^2}{E_0^2+B_0^2}\coth\pqty{\frac{\pi B_0}{E_0}}-\frac{\pi B_0}{E_0}\csch^2\pqty{\frac{\pi B_0}{E_0}}}\sin^2\theta_k\pqty{\sum_{\lambda=\pm}\lambda\frac{k}{\tau}A_{\lambda}}
    .
    \label{A-EoM fullRHS}
}
On the right hand side, both $A_+$ and $A_-$ appear, because of the non-linearity brought by the induced current.
As we saw in Sec.~\ref{The U(1) gauge field without charged particles}, only $A_+$ is amplified by the tachyonic instability for a positive $\xi$, and we expect the hierarchy $|A_+|\gg |A_-|$ remains even in the present case with the induced current.
Thus we drop the contributions from $A_-$ on the right hand side in Eq.~\eqref{A+EoM fullRHS} as sub-leading effects.
Then Eq.~\eqref{A+EoM fullRHS} reads Eq.~\eqref{EoMforA+w/sigma} straightforwardly. 
On the other hand, we cannot ignore the terms with $A_-$ on the right hand side of Eq.~\eqref{A-EoM fullRHS},
because they may be comparable to the left hand side terms.  Eq.~\eqref{A-EoM fullRHS} is rewritten as
\begin{align}
\left[ \partial_\tau^2-\frac{\Sigma_E+\Sigma_{E^{\prime}}\sin^2\theta_{\bm k}}{\tau} \partial_\tau +k^2 - \frac{k}{\tau}\left(2\xi-\Sigma_B-\Sigma_{B^{\prime}}\sin^2\theta_{\bm k} \right) \right] A_-
=
\sin^2\theta_{\bm k}\left(\Sigma_{E^{\prime}} \frac{\partial_\tau}{\tau}+\Sigma_{B^{\prime}}\frac{k}{\tau}\right)A_+.
\end{align}
This equation implies that $A_-$ is sourced by $A_+$.
When we quantized $A_\pm(\tau, \bm k)$ in Eq.~\eqref{quantization}, $\hat{A}_-$ carries
only the creation/annihilation operators for the minus mode, $\hat{a}_{\bm{k}}^{(-)}$ and $\hat{a}_{-\bm{k}}^{(-) \dag}$.
However, to take into account the source effect from $A_+$, we generalize it so that $A_-$
contains the creation/annihilation operators for the plus mode,
\begin{equation}
\hat{A}_-
= \hat{a}_{\bm{k}}^{(-)} \mcA_-^{(\mathrm{int})}  + \hat{a}_{-\bm{k}}^{(-) \dag} \mcA_-^{(\mathrm{int})*}
+\hat{a}_{\bm{k}}^{(+)} \mcA_-^{(\mathrm{src})}  + \hat{a}_{-\bm{k}}^{(+) \dag} \mcA_-^{(\mathrm{src})*},    
\end{equation}
where the mode functions $\mcA_-^{(\mathrm{int})}$ and $\mcA_-^{(\mathrm{src})}$ denote
the intrinsic (homogeneous) solution and the sourced (inhomogeneous) solution, respectively.
We know $\mcA_-^{(\mathrm{int})}$ is negligibly small and hence focus on $\mcA_-^{(\mathrm{src})}$.
Then one obtains 
\begin{align}
    \left[ \partial_\tau^2-\frac{\Sigma_E+\Sigma_{E^{\prime}}\sin^2\theta_{\bm k}}{\tau} \partial_\tau +k^2 - \frac{k}{\tau}\left(2\xi-(\Sigma_B+\Sigma_{B^{\prime}} \sin^2\theta_{\bm k}) \right) \right] \mcA_-^{(\sigma)}(\tau,\bm k)&=
    \sin^2\theta_{\bm k}\left(\Sigma_{E^{\prime}} \frac{\partial_\tau}{\tau}+\Sigma_{B^{\prime}}\frac{k}{\tau}\right)\mcA_+^{(\sigma)}(\tau,\bm k),
\end{align}
where $\mcA_-^{(\sigma)}$ denotes $\mcA_-^{(\mathrm{src})}$,
and $\mcA_+^{(\sigma)}$ is the solution of Eq.~\eqref{EoMforA+w/sigma}.

\bibliographystyle{JHEP}
\bibliography{Paper}
\end{document}